# Valence photoelectron spectra of thiouracils in the gas phase

Dennis Mayer[1,*], Evgenii Titov[2,*], Fabiano Lever[1], Lisa Mehner[3], Marta L. Murillo-Sánchez[4], Constantin Walz[3], John Bozek[5], Peter Saalfrank[2,3], Markus Gühr[1,6]

1) Deutsches Elektronen-Synchrotron DESY, Notkestr. 85, 22607 Hamburg, Germany
2) Institut für Chemie, Universität Potsdam, Karl-Liebknecht-Str. 24/25, 14476 Potsdam, Germany
3) Institut für Physik und Astronomie, Universität Potsdam, Karl-Liebknecht-Str. 24/25, 14476 Potsdam, Germany
4) Max-Planck-Institut für Kernphysik, Saupfercheckweg 1, 69117 Heidelberg, Germany
5) Synchrotron SOLEIL, L'Orme de Merisiers, Départementale 128, 91190 Saint-Aubin, France
6) Institut für Physikalische Chemie, Universität Hamburg, Grindelallee 117, 20146 Hamburg, Germany
* authors to whom the correspondence should be addressed: dennis.mayer@desy.de, titov@uni-potsdam.de

## Abstract

We present a combined experimental and theoretical study of the vibrationally resolved valence photoelectron spectra of the complete series of thiouracils (2-thiouracil, 4-thiouracil and 2,4-dithiouracil) for binding energies between 8 and 17 eV. The theoretical spectra were calculated using equation-of-motion coupled cluster theory for ionization potential (EOM-IP-CCSD) combined with the time-independent double-harmonic adiabatic Hessian approach. For all three thiouracils, the first ionization potential is found between 8.4 and 8.7 eV, which is 1 eV lower than for the canonical nucleobase uracil. Ionization bands up to 12 eV show strong vibrational progressions and are well reproduced by the calculations. These bands are attributed to the ionization of (primarily) sulfur- and oxygen-localized valence molecular orbitals. For higher binding energies, the calculations indicate that nonadiabatic couplings are important for the interpretation of the photoelectron spectra.

## Introduction

Thionated nucleobases have a long history as medical drugs, e.g. as immunosuppressants or photosensitizers in photodynamic therapy.[1,2] This is due to their similarity to canonical nucleobases allowing for substitution in organic tissue and their significantly altered behavior upon excitation with ultraviolet light. The substitution of oxygen by sulfur increases the spin-orbit coupling in the molecules leading to an increased population of long-lived triplet states.[1,2] In these states, radical formation or dimerization may occur, potentially altering DNA strands and their behavior, which can lead to cell death.[3,4]

In recent years, the photodynamics of thiouracils as one of the simplest thionucleobase classes gained considerable attention. Different spectroscopy techniques have been employed to unravel the relaxation pathway towards the triplet states ranging from ultrafast transient absorption[5–9] to ion and electron spectroscopy from the UV to soft x-rays.[10–15] In particular, the charged particle experiments, which were conducted in the gas phase, were often referenced to solution-phase spectra or didn't have well-resolved ground-state reference spectra to begin with. Recently, some authors have started to close these gaps for thiouracils by measuring gas-phase references for the UV and x-ray absorption as well as x-ray photoelectron spectra which were combined with state-of-the-art quantum chemical calculations.[16,17] In the light of time-resolved valence photoelectron spectra performed in the group of S. Ullrich[9,11,12] and simulations of these spectra by Ruckenbauer et al.,[18] we identify a lack of a complete set of high-resolution reference valence photoelectron reference spectra.

In this contribution, we present the static VUV photoelectron spectra of the three molecules, 2- and 4-thiouracil (2- & 4-TU) as well as 2,4-dithiouracil (2,4-dTU), measured at the PLEIADES beamline of the synchrotron SOLEIL. The experimental spectra are discussed with the help of coupled cluster calculations in order to understand the origins of the individual bands as well as observed vibrational progressions. Both experiment and calculations are also compared to previous studies on the molecules and studies on their canonical counterpart uracil.

# Methods

## Experimental Details

The experiment was performed at the PLEIADES beamline of synchrotron SOLEIL.[19] For sample delivery a heatable gas cell was used inside the main vacuum chamber. The samples were heated up to 120°C to achieve sufficient vapor pressure (ca. $1 \cdot 10^{-4}$ mbar).[20] The main chamber pressure was kept at $1 \cdot 10^{-7}$ mbar by means of differential pumping. The pressure inside the gas cell could not be measured but was expected to be between the main chamber pressure and the vapor pressure of the sample.

An Apple II HU 80 permanent magnet undulator was used to generate the VUV beam and a 600 lines/mm grating was employed to select the final photon energy of 100 eV. The beam can enter and exit the gas cell through 10 mm long differential pumping tubes of 2 mm diameter. The focused beam size was about 100 μm x 75 μm where the vertical dimension is typically an image of the monochromator slit. The flux at the given photon energy was on the order of $1.2 \cdot 10^{13}$ photons/s for a 75 μm monochromator slit. The monochromator slit size corresponds to a resolution of ca. 20 meV at a photon energy of 100eV.

Electrons generated by the photon beam exited the gas cell through a slit facing the entrance of the hemispherical electron kinetic energy analyzer (Scienta R4000). The electron spectra were calibrated using a water signal from an initially hydrated sample. The pass energy was chosen such that the overall resolution was estimated to be 30 meV for the measurements.

## Computational Details

Geometry optimizations of the ground state of the neutrals were done at the coupled cluster singles and doubles (CCSD)[21] level, whereas equation-of-motion CCSD for ionization potential (EOM-IP-CCSD)[22,23] was used for the ground and excited states of the cations. No symmetry restrictions were applied. First, the polarized split-valence 6-31G* basis set[24–26] was employed in these calculations. The located minimum energy structures were reoptimized using the triple-zeta cc-pVTZ basis set.[27,28] The vibrational frequencies were calculated numerically using the 6-31G* basis set at the (EOM-IP-)CCSD/6-31G* optimized geometries. The calculations were performed with the Q-Chem 4.4 program.[29]

The adiabatic ionization energies were calculated at the (EOM-IP-)CCSD/cc-pVTZ level and further corrected with zero point energies obtained at the (EOM-IP-)CCSD/6-31G* level to obtain the 0-0 energies (energy differences between lowest vibrational states of $S_0$ and $D_i$ electronic states). We note that test calculations demonstrated very minor differences when using the cc-pVDZ frequencies instead of the 6-31G* frequencies. The EOM-IP-CCSD calculations of the (numerical) frequencies with the cc-pVTZ basis set are too demanding.

The vibrationally resolved photoelectron spectra were computed with the ezSpectrum 3.0 program.[30,31] They were obtained using the time-independent double-harmonic adiabatic Hessian approach.[32] The frequency alteration and Duschinsky rotations were taken into account. Up to two vibrational quanta were included for the initial ($S_0$) state, and up to five in the final ($D_i$) states. The spectra were calculated at the temperature of 120 °C (393.15 K). The intensity threshold was set to $10^{-6}$. The vibronic stick spectra $\{E_j, I_j\}$ were broadened with Lorentzians as

$$I(E) = \sum_j I_j \frac{\gamma^2}{(E - E_j)^2 + \gamma^2}$$

with = 0.02 eV. Here, $E_j$ and $I_j$ are vibronic excitation energies and corresponding intensities, respectively.

The vertical electronic spectra were calculated using EOM-IP-CCSD/cc-pVTZ at the CCSD/cc-pVTZ optimized geometries. The intensities of the transitions were estimated as the product of left and right Dyson orbital norms.[33] The Dyson orbitals (and their norms) were calculated using Q-Chem 4.4.

# Results and Discussion

## Experimental Valence Spectra

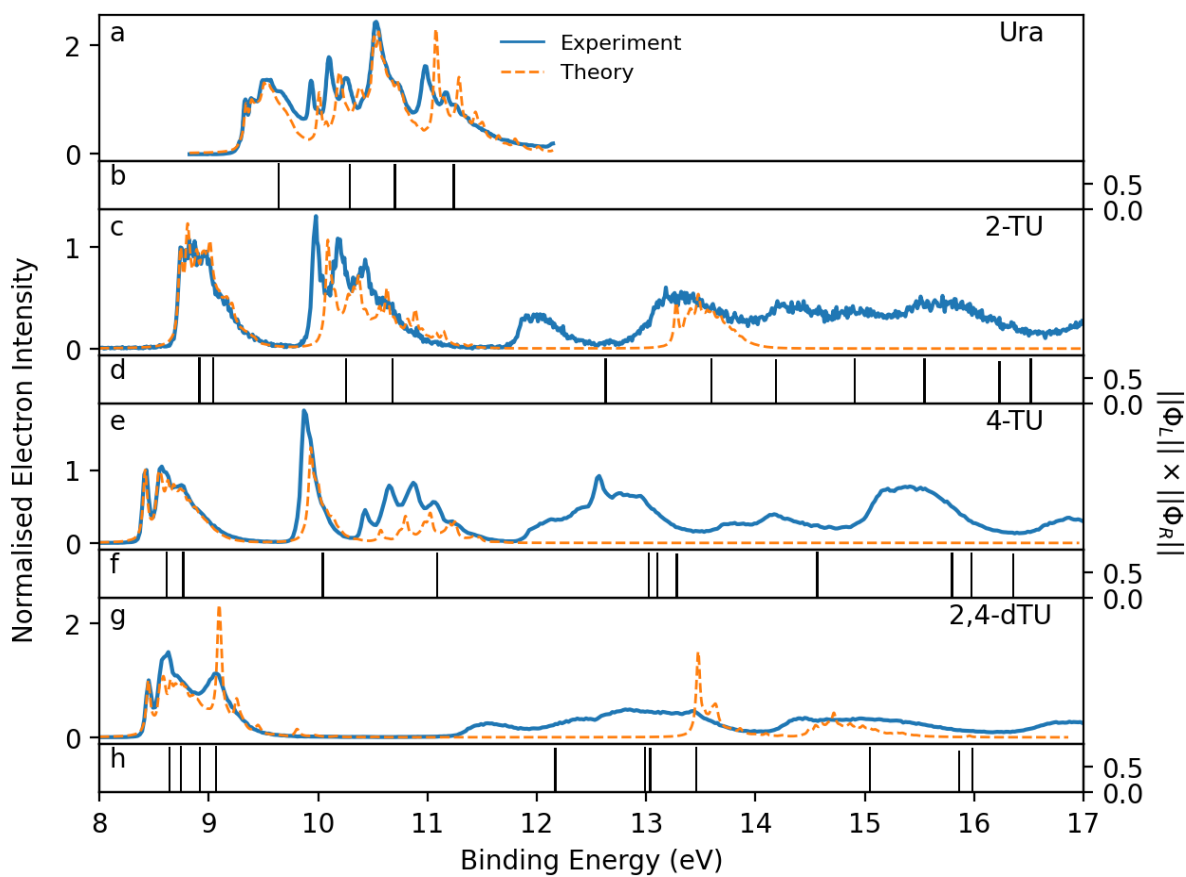

**Figure 1:** Experimental (solid, blue line) taken at hv=100eV together with theoretical (dashed, orange line) photoelectron spectra for uracil (a) and the three thiouracils (c, e, g) together with vertical excitation (stick) spectra (b, d, f, h). Shifts of 0.05 eV, 0.1 eV, -0.02 eV and -0.07 eV were applied to the theoretical spectra of Ura, 2-, 4-TU and 2,4-dTU, respectively, to match the experimental spectra. The experimental uracil spectrum was taken from Fulfer et al.[34] at a photon energy of 40 eV.

We start with the discussion of the experimental spectra. Figure 1 shows the valence photoelectron spectra of 2- and 4-thiouracil (2-TU & 4-TU) as well as 2,4-dithiouracil (2,4-dTU) alongside the theoretical results. For comparison, a uracil spectrum is also shown in panel a, with the experimental spectrum taken from Fulfer et al.[34] No background has been subtracted from the spectra as after careful comparison with reference spectra of usual contaminations (i.e. $N_2$ and $H_2O$) we found that any background should be negligible. The measured electron kinetic energy window (80 - 95eV) was in a range where the transmission function of the hemisphere is flat. Thus, no further normalization was done on the spectra.

The first ionization band has its first maximum at energies of 8.75 eV for 2-TU and 8.42 eV and 8.45 eV for 4-TU and 2,4-dTU, respectively. For 2-TU this energy is in agreement with previous literature.[10,35] In all three molecules, this first band shows some structure indicating vibrational excitations. Compared to the canonical nucleobase uracil, the band is shifted on average by 0.9 eV towards lower binding energies upon thionation.[34,36,37]

**Table 1:** Experimental ionization energies in eV for the three thiouracils and uracil.

|  | **2-TU** | **4-TU** | **2,4-dTU** | **Ura** |
|---|---|---|---|---|
| **This work** | 8.74 ± 0.03 | 8.42 ± 0.01 | 8.45 ± 0.01 | - |
| **Literature** | 8.70[10]<br>8.80[35] | - | - | 9.3[34]<br>9.46[36]<br>9.56[37] |

A second band appears at 9.98 and 9.83 eV for 2-TU and 4-TU, respectively, which are followed by a strong vibrational progression. In the case of 2-TU, the following three identifiable peaks are separated by ca. 0.22 eV. For 4-TU, the vibrationally resolved band reaches from 10.3 eV to 11.5 eV covering ca. 5 peaks separated by ca. 0.2 eV. Interestingly, 2,4-dTU does not show any feature in the range between 9.5 to 11.5 eV. However, as mentioned above, the first valence band of uracil falls into the same energy range. This indicates that the second band might be associated with orbitals localized at the oxygen atom, as the absence of oxygen is what distinguishes 2,4-dTU from the other three molecules. Also, the separation of the peaks in 2- and 4-TU (0.20–0.22 eV or 1600–1800 cm$^{-1}$) is similar to the frequency of the C=O bond stretch. As a consequence, one might assume that the first band in each of the three thionated molecules can be attributed to orbitals localized at the sulfur atom. The assignment to the atoms is verified by our theoretical calculations which are discussed in the following sections.

Beyond 11.5 eV more bands are visible, however, they are not structured as the first two. For 2-TU, a band around 12 eV is observed followed by a broad band between 12.6 and 16.5 eV which shows smaller peaks at ca. 13.1, 14.2, 14.9 and 15.5 eV. 4-TU shows a broader band between 11.9 and 13.2 eV with a distinct peak at ca. 12.5 eV. This band is followed by two smaller peaks around 13.8 and 14.2 eV. Another band ranges between 15 and 16 eV with no distinct peak. The doubly thionated uracil shows a broad band between 11.3 and 13.8 eV with small peaks at around 11.5, 12.4, 12.8 and 13.4 eV. Another band appears between 14.2 and 16 eV with smaller peaks around 14.4 and 16 eV.

# Vertical electronic transitions

The vertical ionization potentials and dominant Hartree–Fock canonical molecular orbitals (and their energies) involved in ionization are shown in Tabs. 2, 3 and 4 for 2-TU, 4-TU and 2,4-dTU, respectively. In agreement with Ruckenbauer et al,[18] we find that the lowest energy band in the vibrationally resolved spectrum stems from transitions solely to $D_0$ and $D_1$ for both 2-TU and 4-TU, whereas for 2,4-dTU it originates from the lowest 4 transitions, to $D_0$–$D_3$ states. The corresponding molecular orbitals are dominated by $n$ or $\pi$ type contributions of sulfur.

Interestingly, for 2-TU, the leading orbital number in column 3 is monotonically decreasing showing that the lowest state ($D_0$) corresponds to ionization from the HOMO (Highest Occupied Molecular Orbital), the next state ($D_1$) to HOMO-1, $D_2$ to HOMO-2, and so on in agreement with the Koopmans' theorem. In contrast, this is different for 4-TU and 2,4-dTU. For example, $D_0$ of 4-TU arises from HOMO-1, whereas $D_1$ from HOMO. The same is observed for 2,4-dTU. We note that in these cases, the energy gaps between both multi-electron states ($D_0$/$D_1$) and orbitals (HOMO/HOMO-1) are small, ~0.1–0.2 eV.

The $D_2$ and $D_3$ states of 2-TU and 4-TU correspond predominantly to ionization from oxygen (again, either from $n$ or $\pi$ type oxygen orbitals, see Tabs. 2, 3) with some contribution of the ring and the sulphur atom. This explains the lack of electronic transitions in the associated energy range (~10–11 eV) for 2,4-dTU which does not contain oxygen.

The first four transitions of all the molecules originate from ionization from four highest molecular orbitals (HOMO to HOMO-3). The higher energy states correspond to ionization from lower orbitals, which often are dominated by a ring contribution but in some cases also involve oxygen and/or sulphur atoms (see Tabs. 2–4).

It is instructive to note that Koopmans' ionization potentials calculated as minus energies of respective leading molecular Hartree–Fock (canonical) orbitals are overestimated by 0.2–2.3 eV with respect to the EOM-IP-CCSD values (compare columns 2 and 4 of Tabs. 2–4). Specifically, the maximal and minimal deviations are 0.36 and 2.11 eV for 2-TU, 0.22 and 2.33 for 4-TU, and 0.30 and 2.19 eV for 2,4-dTU. This clearly shows the limitation of a simple one-electron picture based on HF orbitals.

We have also calculated the Dyson orbitals for the $S_0$–$D_i$ transitions. As can be seen from Tabs. 2–4, they are (very) similar to the corresponding canonical Hartree–Fock orbitals for all considered transitions (compare two last columns of Tabs. 2–4).

Finally, we note that the Dyson orbital norms (presented in Fig. 1) are all rather large (left and right norms separately >0.9, and their products >0.8) for the lowest eleven EOM-IP-CCSD transitions, for all thionated molecules. This is in contrast to Ruckenbauer et al.[18], where some transitions possess close-to-zero Dyson orbital norms (see the supporting information of ref. [18]). We attribute this discrepancy to the use of different methods, EOM-IP-CCSD here vs. MRCIS in ref. [18].

## Vibronic transitions

To calculate vibronic transitions we applied the time-independent adiabatic Hessian approach which requires geometry optimizations for both initial ($S_0$ in our case) and final ($D_i$ in our case) states and the normal modes and vibrational frequencies for the both states. We attempted to optimize $D_0$–$D_{10}$ states for each thionated molecule using EOM-IP-CCSD (and $S_0$ using CCSD). For uracil, only states $D_0$–$D_3$ were optimized corresponding to the energy range of the experimental spectrum of Fig. 1a.

For all thionated molecules, we were able to optimize $D_0$–$D_4$ (and $S_0$) states with the exception of $D_2$ for 2,4-dTU. Optimizations of higher lying cationic states often led to crossing regions (with a lower energy state) instead of zero-gradient minima of the states being optimized. This was also observed in the $D_2$ optimizations for 2,4-TU. However, the minima of the following higher energy states were located: $D_5$ for 2-TU, $D_8$ for 4-TU, and $D_7$ and $D_8$ for 2,4-dTU.

For the successfully optimized states, the vibrationally resolved spectra were calculated as Franck–Condon factors weighted with temperature-dependent coefficients according to the Boltzmann distribution. The stick spectra were concatenated and broadened with Lorentzians (see Methods) to obtain the total vibrationally resolved spectra shown in Fig. 1.

The vibronic stick spectra for the transitions to $D_0$–$D_3$ are shown in Fig. 2 and assignment of the major vibronic transitions are provided in the Appendix A. We note that the time-independent approach to calculate vibronic spectra (used in this work) provides direct information on which normal modes contribute to which transitions. The 0-0 transitions are the most intense almost for all considered cases. However, for the $D_3$ state of 4-TU the most intense stick transition corresponds to excitation of mode 26 (NH bending + CO stretch + CC stretch, see Appendix A). For the $D_3$ state of 2-TU, this transition (excitation of mode 26) is similar in intensity to the 0-0 transition. Further excitations of mode 26 (adding more vibrational quanta) are responsible for the observed progression of the $S_0 \rightarrow D_3$ transition in 4-TU and 2-TU (at ~10–11 eV in the experimental spectra of Fig. 1 and 2). The $D_3$ intensity for the 2,4-dTU is largely overestimated compared to the experimental spectrum.

The $D_4$ optimized geometries are considerably non-planar (in contrast to the $D_0$–$D_3$ geometries), and the calculated Franck–Condon factors are very small for the $S_0 \rightarrow D_4$ transitions of all three molecules (see Appendix B). The same was also observed for the $D_8$ state of 4-TU.

The optimized geometries of the $D_5$ state of 2-TU and the $D_7$ and $D_8$ states of 2,4-dTU are, in turn, planar and the resulting Franck–Condon factors are relatively large. At that, the $D_7$ intensity is strongly overestimated in comparison to experiment (Fig. 1).

Finally, we note that inclusion of nonadiabatic couplings (multistate treatment) seems to be necessary to properly describe the higher energy part of the spectra,[38,39] because the undertaken geometry optimizations for higher energy cationic excited states usually hit intersection regions suggesting importance of the latter. To account for the nonadiabatic effects, one could attempt to construct a vibronic Hamiltonian and either diagonalize it or perform quantum dynamics simulations to eventually calculate the photoelectron spectra[38,39,40] (which goes beyond the double-harmonic adiabatic Hessian approach). However, this is beyond the scope of the present work.

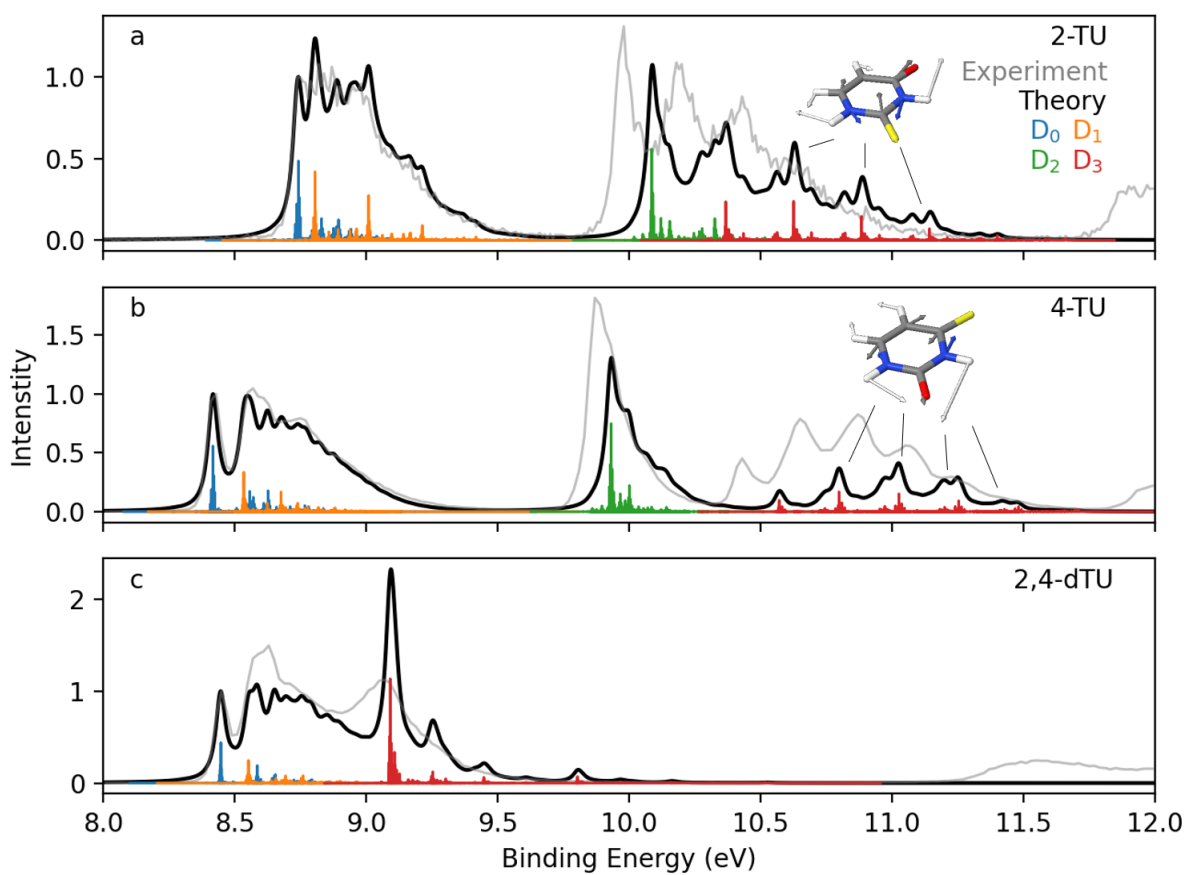

**Figure 2:** Vibrationally-resolved spectra between 8 and 12 eV. The contributions of the different cationic states to the theoretical spectrum (black) are colored ($D_0$ - blue, $D_1$ - orange, $D_2$ - green, $D_3$ - red). The insets for 2- and 4-TU show the most prominent vibrational mode for the $S_0 \rightarrow D_3$ transition. The other vibronic transitions contributing to the spectrum are shown in the Appendix. The experimental spectra are shown in gray. The same shifts as in Figure 1 have been applied to the theoretical spectra.

**Table 2:** Vertical ionization potentials (calculated using EOM-IP-CCSD/cc-pVTZ and Koopmans' theorem) and molecular orbitals (canonical and Dyson) involved in ionization for the lowest eleven transitions of 2-TU. Orbital energies refer to Hartree–Fock. The HOMO is orbital no. 33.

| Transition | Excitation energy (eV) | Leading orbital (contribution) | Minus orbital energy (eV) | Picture of the canonical orbital | Picture of the Dyson orbital |
|---|---|---|---|---|---|
| $S_0 \to D_0$ | 8.814 | 33 (0.93) | 9.170 | 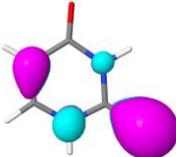 | 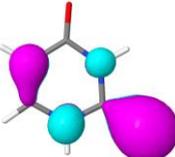 |
| $S_0 \to D_1$ | 8.938 | 32 (0.93) | 9.497 | 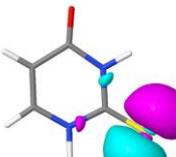 | 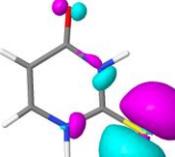 |
| $S_0 \to D_2$ | 10.158 | 31 (0.88) | 10.857 | 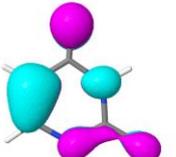 | 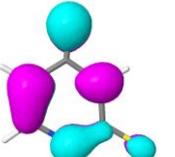 |
| $S_0 \to D_3$ | 10.579 | 30 (0.89) | 12.517 | 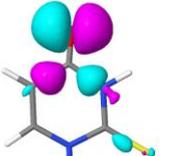 | 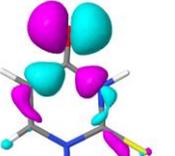 |
| $S_0 \to D_4$ | 12.528 | 29 (0.86) | 13.551 | 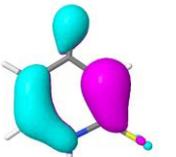 | 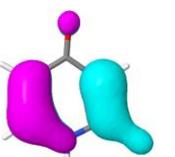 |
| S$_0$→D$_5$ | 13.499 | 28 (0.90) | 14.803 | 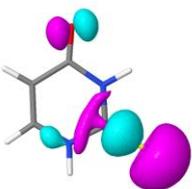 | 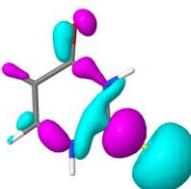 |
| S$_0$→D$_6$ | 14.084 | 27 (0.88) | 15.592 | 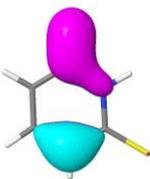 | 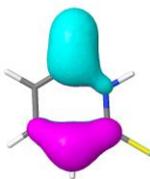 |
| S$_0$→D$_7$ | 14.807 | 26 (0.80) | 16.653 | 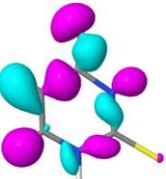 | 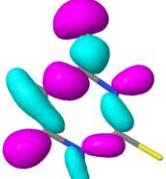 |
| S$_0$→D$_8$ | 15.445 | 25 (0.78) | 17.143 | 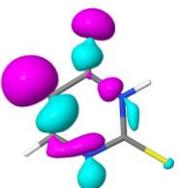 | 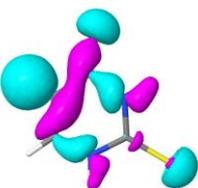 |
| S$_0$→D$_9$ | 16.132 | 24 (0.86) | 18.096 | 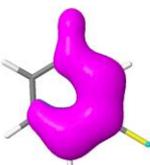 | 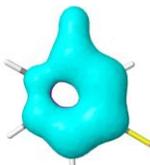 |
| S$_0$→D$_{10}$ | 16.417 | 23 (0.84) | 18.531 | 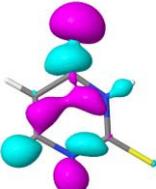 | 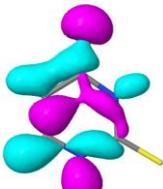 |

**Table 3:** Vertical ionization potentials (calculated using EOM-IP-CCSD/cc-pVTZ and Koopmans' theorem) and molecular orbitals (canonical and Dyson) involved in ionization for the lowest eleven transitions of 4-TU. The HOMO is orbital 33.

| Transition | Excitation energy (eV) | Leading orbital (contribution) | Minus orbital energy (eV) | Picture of the canonical orbital | Picture of the Dyson orbital |
|---|---|---|---|---|---|
| $S_0 \rightarrow D_0$ | 8.635 | 32 (0.93) | 9.225 | 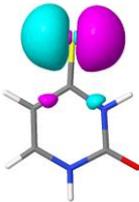 | 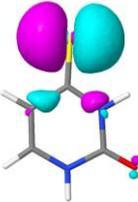 |
| $S_0 \rightarrow D_1$ | 8.786 | 33 (0.93) | 9.007 | 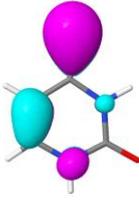 | 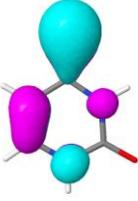 |
| $S_0 \rightarrow D_2$ | 10.062 | 31 (0.91) | 10.857 | 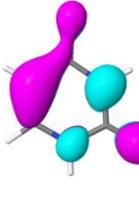 | 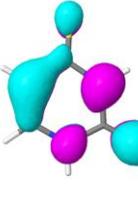 |
| $S_0 \rightarrow D_3$ | 11.106 | 30 (0.88) | 13.007 | 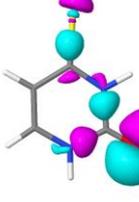 | 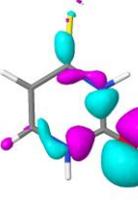 |
| $S_0 \rightarrow D_4$ | 13.043 | 27 (0.83) | 14.613 | 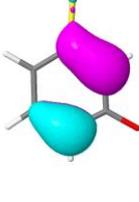 | 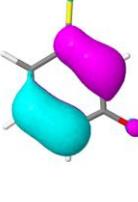 |
| S$_0$→D$_5$ | 13.119 | 28 (0.88) | 14.449 | 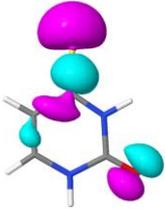 | 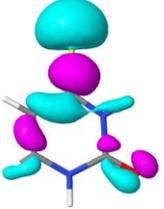 |
| S$_0$→D$_6$ | 13.299 | 29 (0.80) | 14.395 | 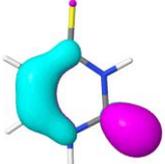 | 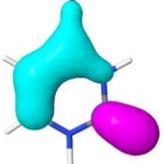 |
| S$_0$→D$_7$ | 14.585 | 26 (0.90) | 16.218 | 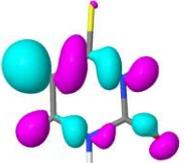 | 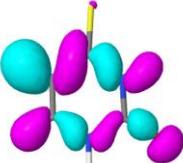 |
| S$_0$→D$_8$ | 15.818 | 24 (0.66) | 18.150 | 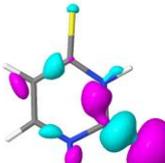 | 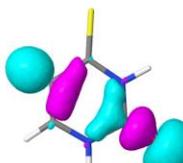 |
| S$_0$→D$_9$ | 15.997 | 25 (0.66) | 17.715 | 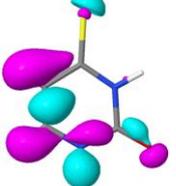 | 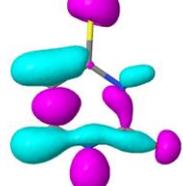 |
| S$_0$→D$_{10}$ | 16.376 | 23 (0.87) | 18.368 | 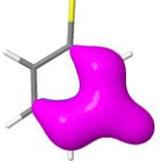 | 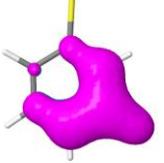 |

**Table 4:** Vertical ionization potentials (calculated using EOM-IP-CCSD/cc-pVTZ and Koopmans' theorem) and molecular orbitals (canonical and Dyson) involved in ionization for the lowest eleven transitions of 2,4-dTU. The HOMO is orbital 37.

| Transition | Excitation energy (eV) | Leading orbital (contribution) | Minus orbital energy (eV) | Picture of the canonical orbital | Picture of the Dyson orbital |
|---|---|---|---|---|---|
| $S_0 \rightarrow D_0$ | 8.713 | 36 (0.92) | 9.361 | 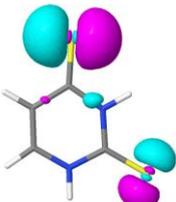 | 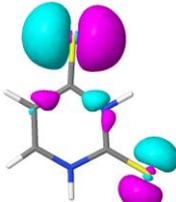 |
| $S_0 \rightarrow D_1$ | 8.815 | 37 (0.86) | 9.116 | 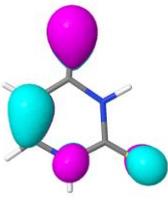 | 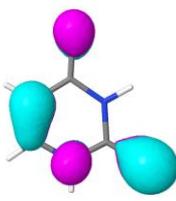 |
| $S_0 \rightarrow D_2$ | 8.989 | 35 (0.84) | 9.442 | 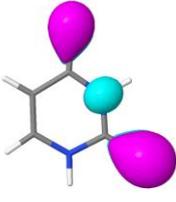 | 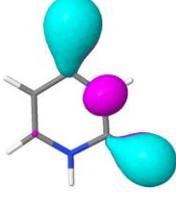 |
| $S_0 \rightarrow D_3$ | 9.137 | 34 (0.92) | 9.769 | 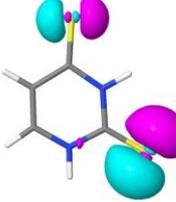 | 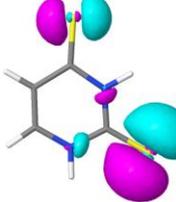 |
| $S_0 \rightarrow D_4$ | 12.238 | 33 (0.90) | 12.871 | 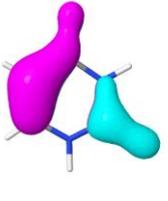 | 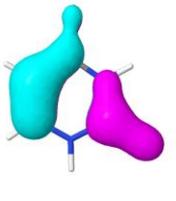 |
| S$_0$→D$_5$ | 13.062 | 32 (0.91) | 14.313 | 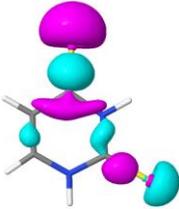 | 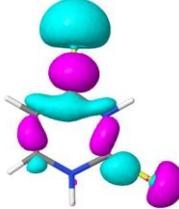 |
| S$_0$→D$_6$ | 13.107 | 31 (0.88) | 14.776 | 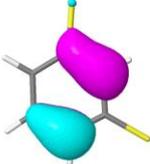 | 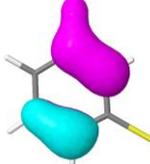 |
| S$_0$→D$_7$ | 13.526 | 30 (0.90) | 14.857 | 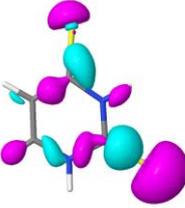 | 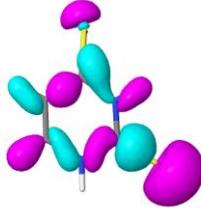 |
| S$_0$→D$_8$ | 15.116 | 29 (0.90) | 16.817 | 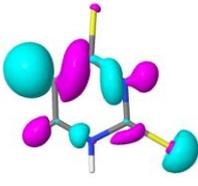 | 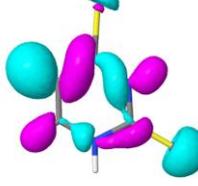 |
| S$_0$→D$_9$ | 15.935 | 27 (0.84) | 18.123 | 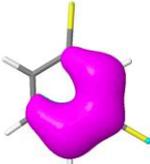 | 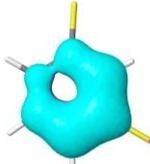 |
| S$_0$→D$_{10}$ | 16.052 | 28 (0.89) | 17.932 | 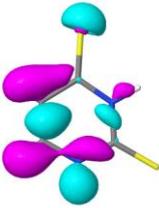 | 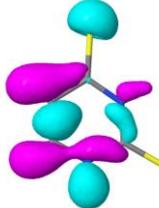 |

# Conclusion

In this study, we reported the vibrationally-resolved gas-phase valence photoelectron spectra of three thiouracils up to 17 eV. The data for 2-TU is in agreement with previously published results, the spectra for 4-TU and 2,4-dTU are presented for the first time to the best of our knowledge. The direct comparison of the three spectra in combination with the photoelectron spectrum of uracil allowed to assign the first emerging band to the ionization of sulfur-dominated orbitals and the second band in 2- and 4-TU to oxygen-dominated orbitals. EOM-IP-CCSD calculations confirmed the assignment. The calculations were further combined with the time-independent double-harmonic adiabatic Hessian approach to simulate the vibrational progression of the ionization band. This resulted in a good match between theory and experiment for the first two ionization bands and allowed insight into the vibrational modes visible in the spectra. However, this approach fails to reproduce the bands for higher cationic states of the molecules. This is likely due to a missing treatment of nonadiabatic couplings between higher energy states.

# Acknowledgements

The spectra were recorded at the PLEIADES beamline at the synchrotron SOLEIL, France, under proposals 20200549 and 20211636. We acknowledge SOLEIL for provision of synchrotron radiation facilities and we would like to thank E. Robert and C. Nicolas for their technical assistance during the beamtime at PLEIADES. We thank BMBF for funding via Verbundforschungsprojekt 05K19IP1. We acknowledge DFG funding via grants GU 1478/1-1 (M. G.) and SA 547/17-1 (P. S.) via the common project 445713302. We thank the Deutsche Forschungsgemeinschaft (DFG, German Research Foundation) for financial support via CRC/SFB 1636 – Project ID 510943930 – Project Nos. A03 & B05.

# Author Declarations

## Conflict of interest

The authors have no conflict of interest.

# Author Contributions

**Dennis Mayer:** Data curation (equal); Formal analysis (equal); Investigation (equal); Validation (equal); Visualization (equal); Writing – original draft (equal); Writing – review & editing (equal). **Evgenii Titov**: Data curation (equal); Formal analysis (equal); Investigation (equal); Methodology (lead); Software (lead); Validation (equal); Visualization (equal); Writing – original draft (equal); Writing – review &

editing (equal). **Fabiano Lever:** Investigation (equal); Writing – review & editing (equal). **Lisa Mehner:** Investigation (equal); Writing – review & editing (equal). **Marta L. Murillo-Sanchez:** Investigation (equal); Writing – review & editing (equal). **Constantin Walz:** Investigation (equal); Writing – review & editing (equal). **John Bozek:** Investigation (equal); Resources (equal); Writing – review & editing (equal). **Peter Saalfrank:** Funding acquisition (equal); Project administration (equal); Resources (equal); Supervision (equal); Writing – review & editing (equal). **Markus Gühr:** Conceptualization (lead); Funding acquisition (equal); Investigation (equal); Project administration (equal); Supervision (equal); Writing – review & editing (equal).

## Data Availability

The experimental raw data were generated at the Synchrotron SOLEIL. Derived data and the simulated results supporting the findings of the study are openly available in Zenodo at https://www.doi.org/10.5281/zenodo.15720936.

# Appendix A. Assignment of major vibronic transitions

2-TU

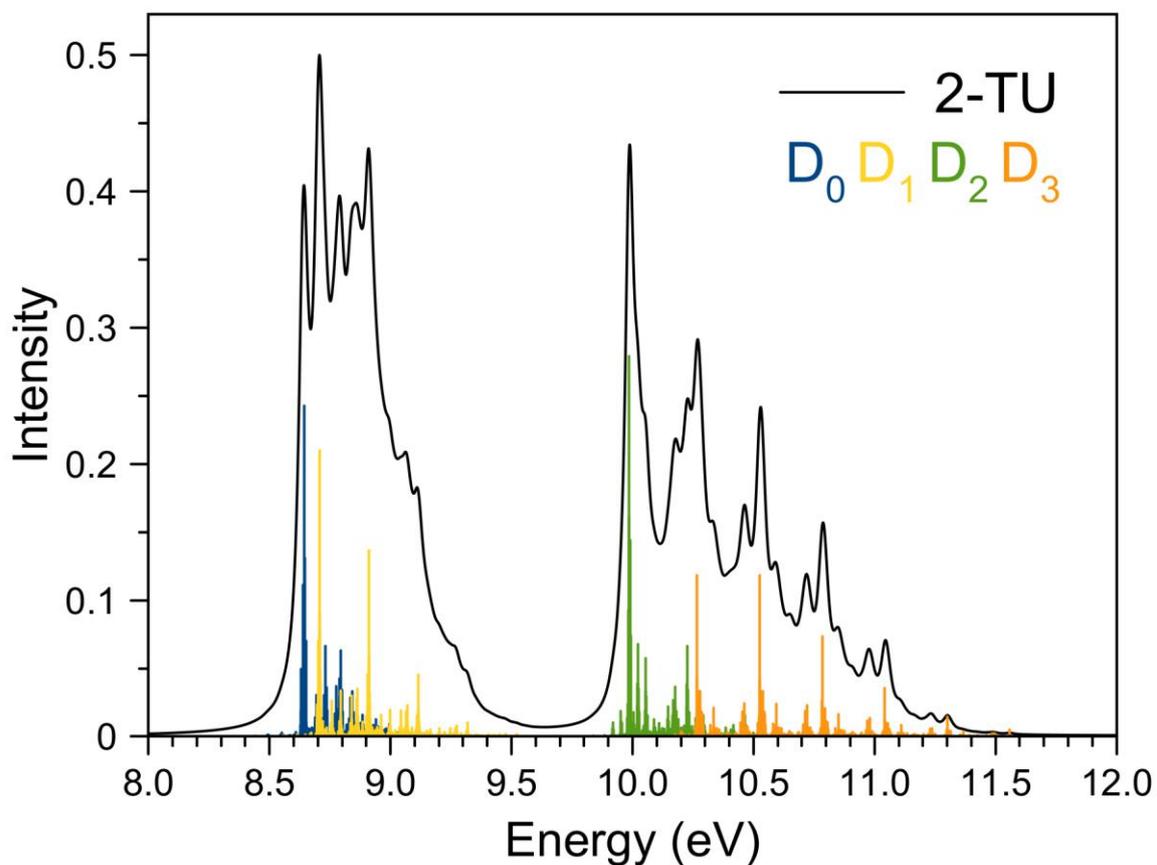

**Figure A1:** Vibronic stick spectra for the lowest four transitions (from $S_0$ to $D_0$–$D_3$) and the broadened spectrum for 2-TU.

**Table A1:** The $S_0$ normal modes of 2-TU involved in the major vibronic transitions from $S_0$ to $D_0$–$D_3$.

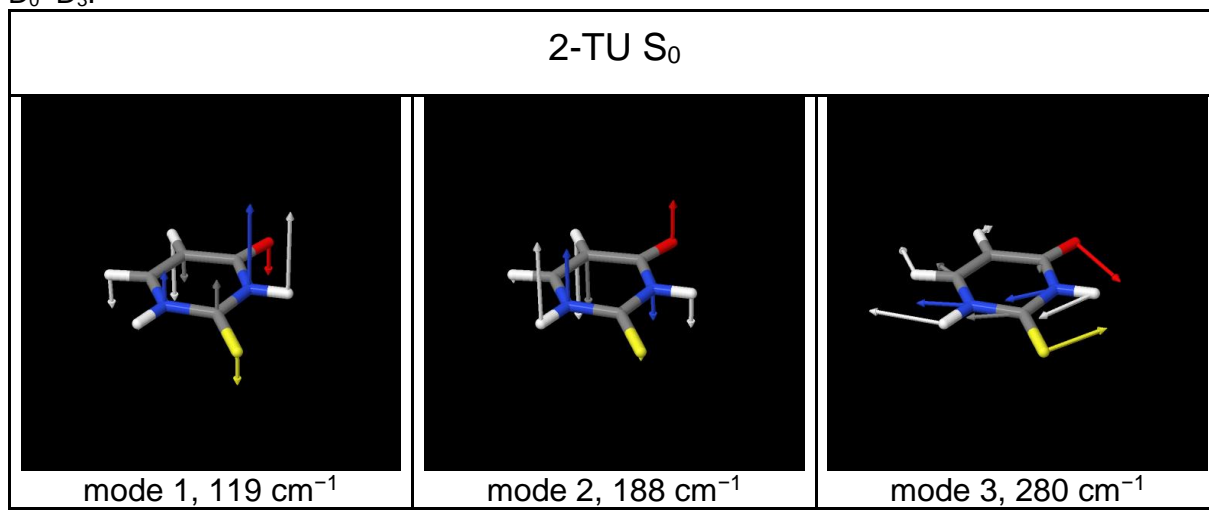

$S_0 \rightarrow D_0$

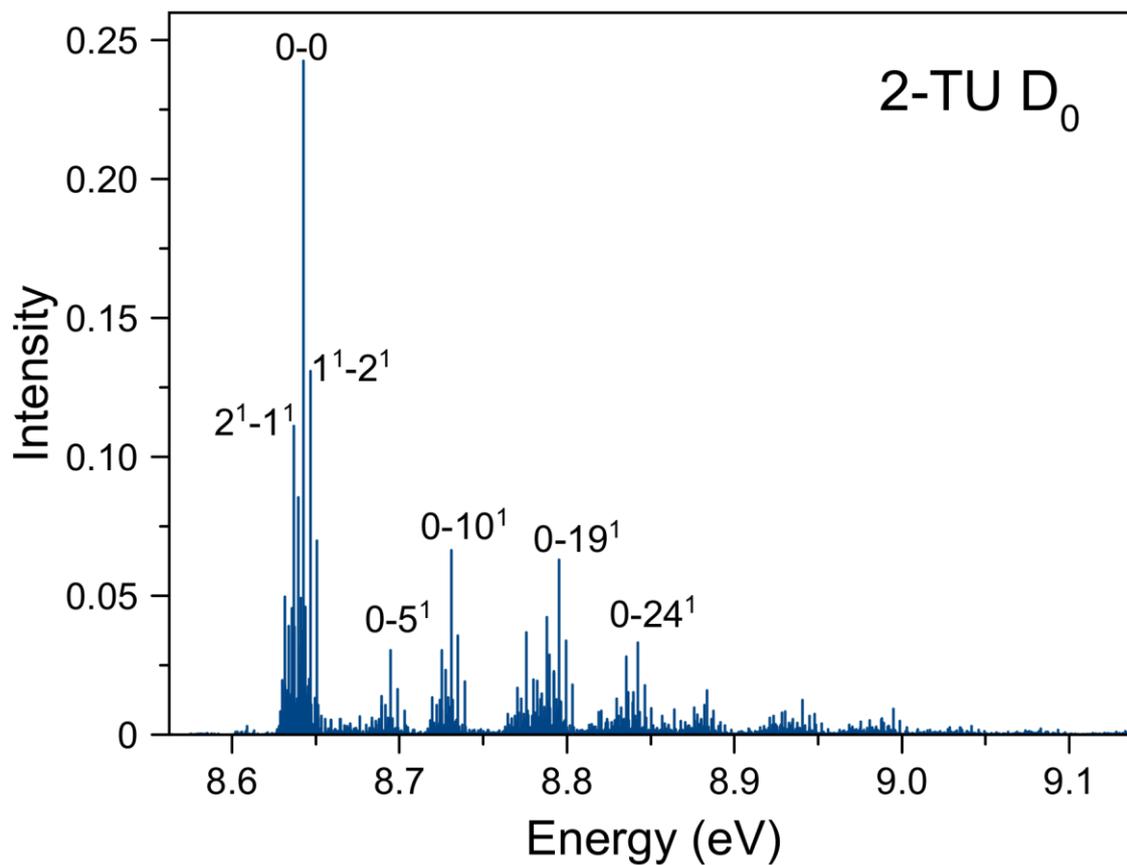

**Figure A2:** $S_0 \rightarrow D_0$ stick spectrum of 2-TU with assignment of major transitions. The used notation is $g^m$-$e^n$, where $g$ is the excited normal mode in the $S_0$ state, $e$ is the excited normal mode in the $D_0$ state, and $m$ and $n$ are the numbers of vibrational quanta. "0" stands for the lowest vibrational states of $S_0$ and $D_0$.

**Table A2:** The $D_0$ normal modes of 2-TU involved in the major $S_0 \rightarrow D_0$ vibronic transitions.

| 2-TU $D_0$ | | |
|---|---|---|
| 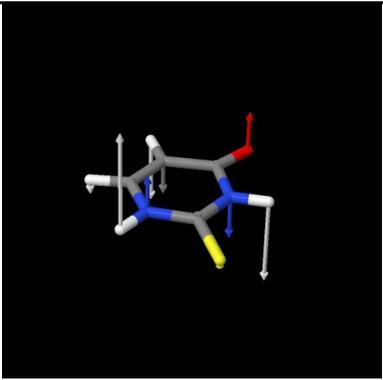 mode 1, 112 cm$^{-1}$ | 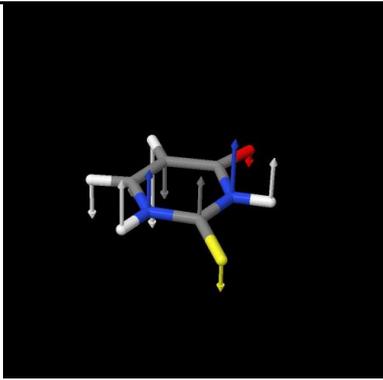 mode 2, 163 cm$^{-1}$ | 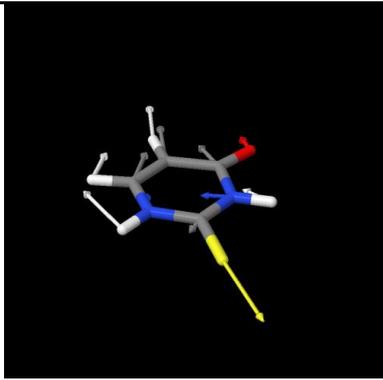 mode 5, 420 cm$^{-1}$ |
| 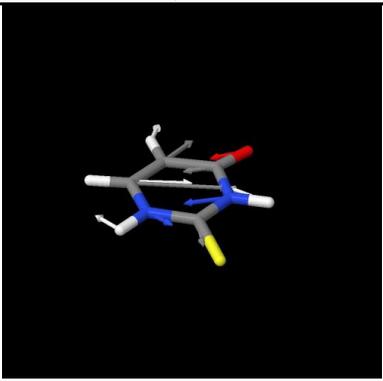 mode 10, 711 cm$^{-1}$ | 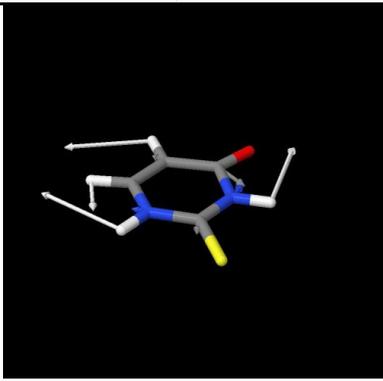 mode 19, 1231 cm$^{-1}$ | 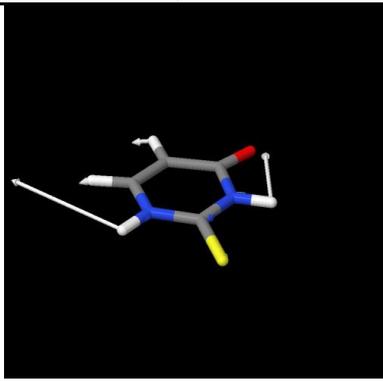 mode 24, 1609 cm$^{-1}$ |

S$_0$→D$_1$

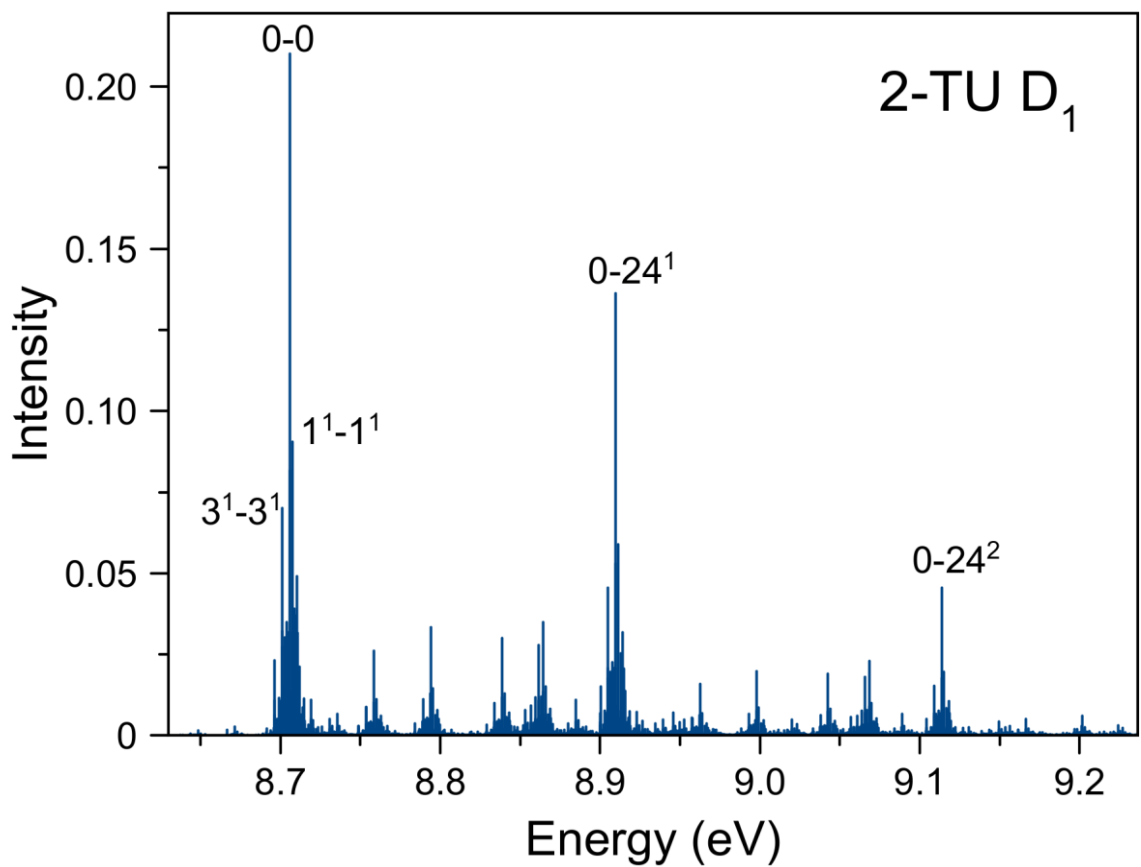

**Figure A3:** S$_0$→D$_1$ stick spectrum of 2-TU with assignment of major transitions. The used notation is $g^m$-$e^n$, where $g$ is the excited normal mode in the S$_0$ state, $e$ is the excited normal mode in the D$_1$ state, and $m$ and $n$ are the numbers of vibrational quanta. "0" stands for the lowest vibrational states of S$_0$ and D$_1$.

**Table A3:** The D$_1$ normal modes of 2-TU involved in the major S$_0$→D$_1$ vibronic transitions.

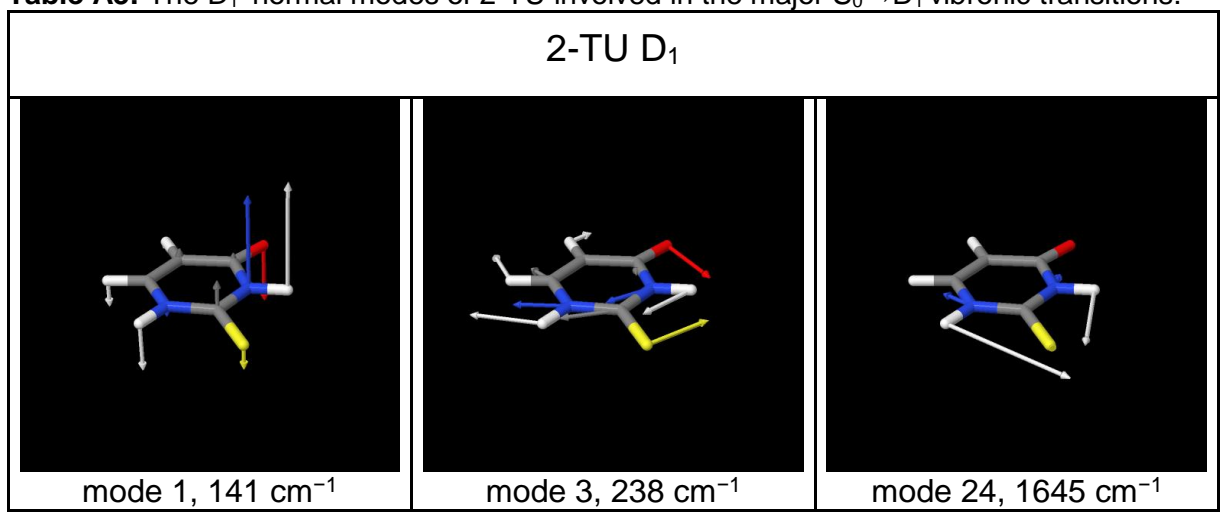

S$_0$→D$_2$

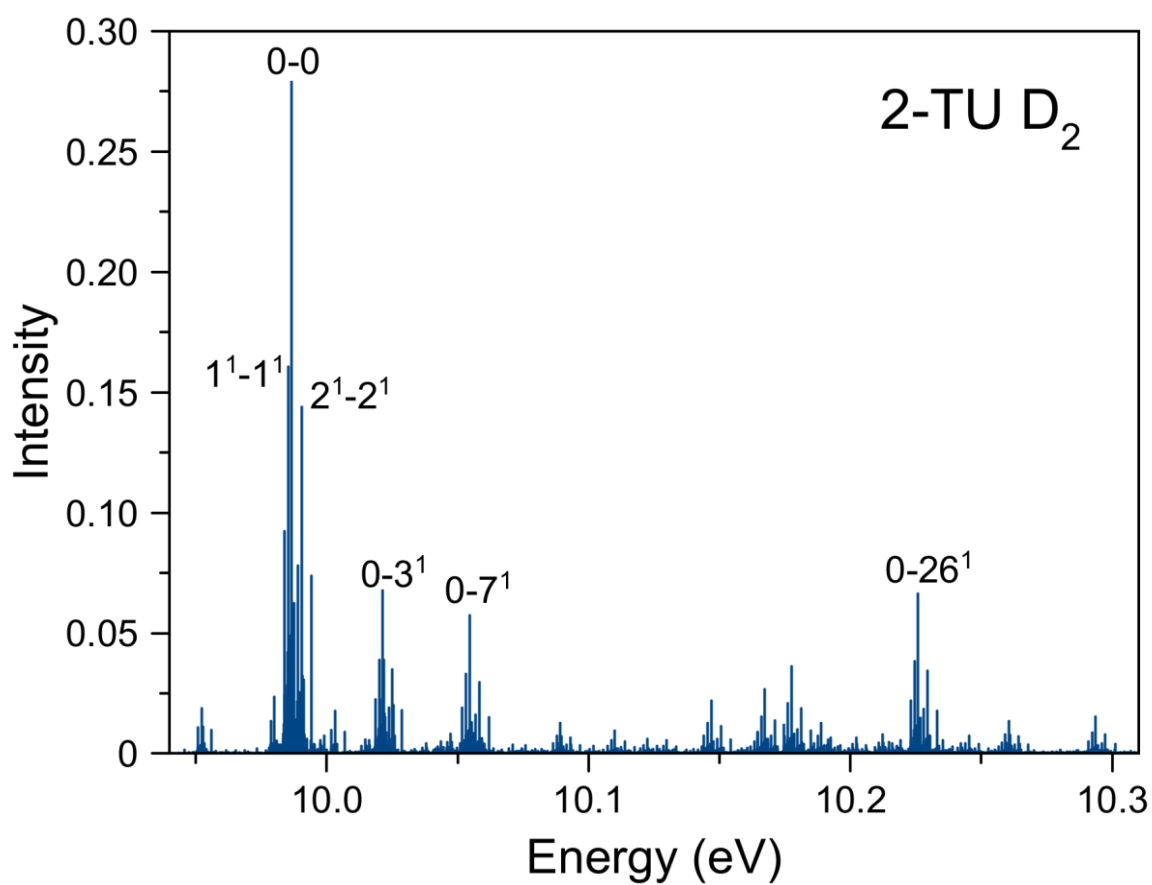

**Figure A4:** S$_0$→D$_2$ stick spectrum of 2-TU with assignment of major transitions. The used notation is $g^m$-$e^n$, where $g$ is the excited normal mode in the S$_0$ state, $e$ is the excited normal mode in the D$_2$ state, and $m$ and $n$ are the numbers of vibrational quanta. "0" stands for the lowest vibrational states of S$_0$ and D$_2$.

**Table A4:** The $D_2$ normal modes of 2-TU involved in the major $S_0 \rightarrow D_2$ vibronic transitions.

| 2-TU $D_2$ | | |
|---|---|---|
| 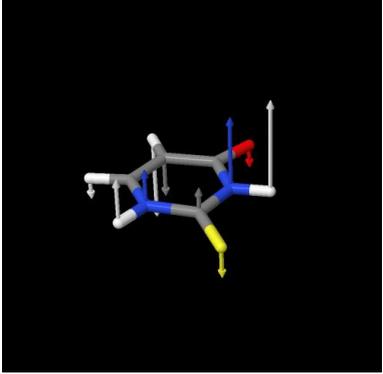 mode 1, 119 cm$^{-1}$ | 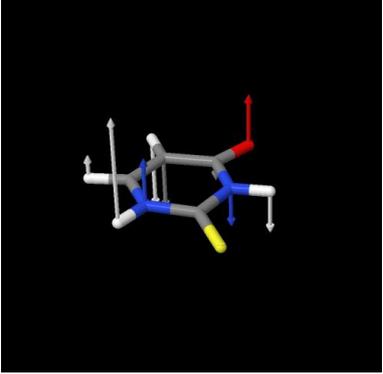 mode 2, 188 cm$^{-1}$ | 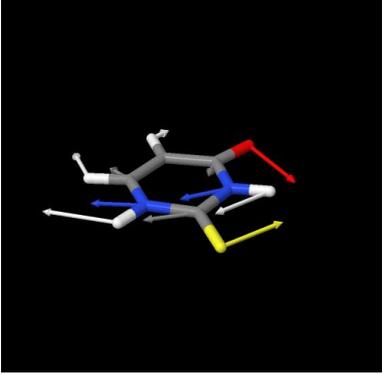 mode 3, 280 cm$^{-1}$ |
| 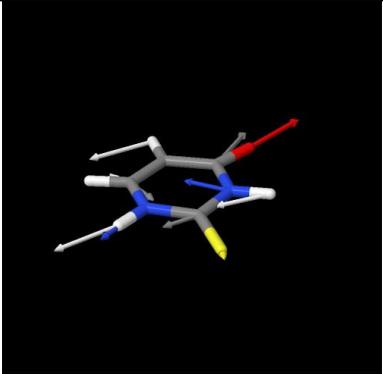 Mode 7, 548 cm$^{-1}$ | 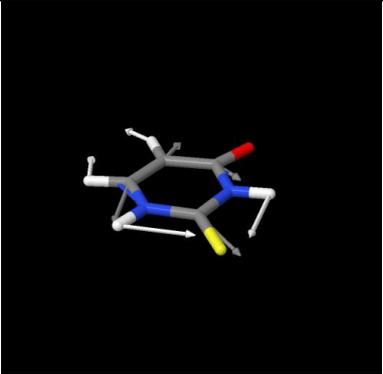 mode 26, 1929 cm$^{-1}$ | |

$S_0 \rightarrow D_3$

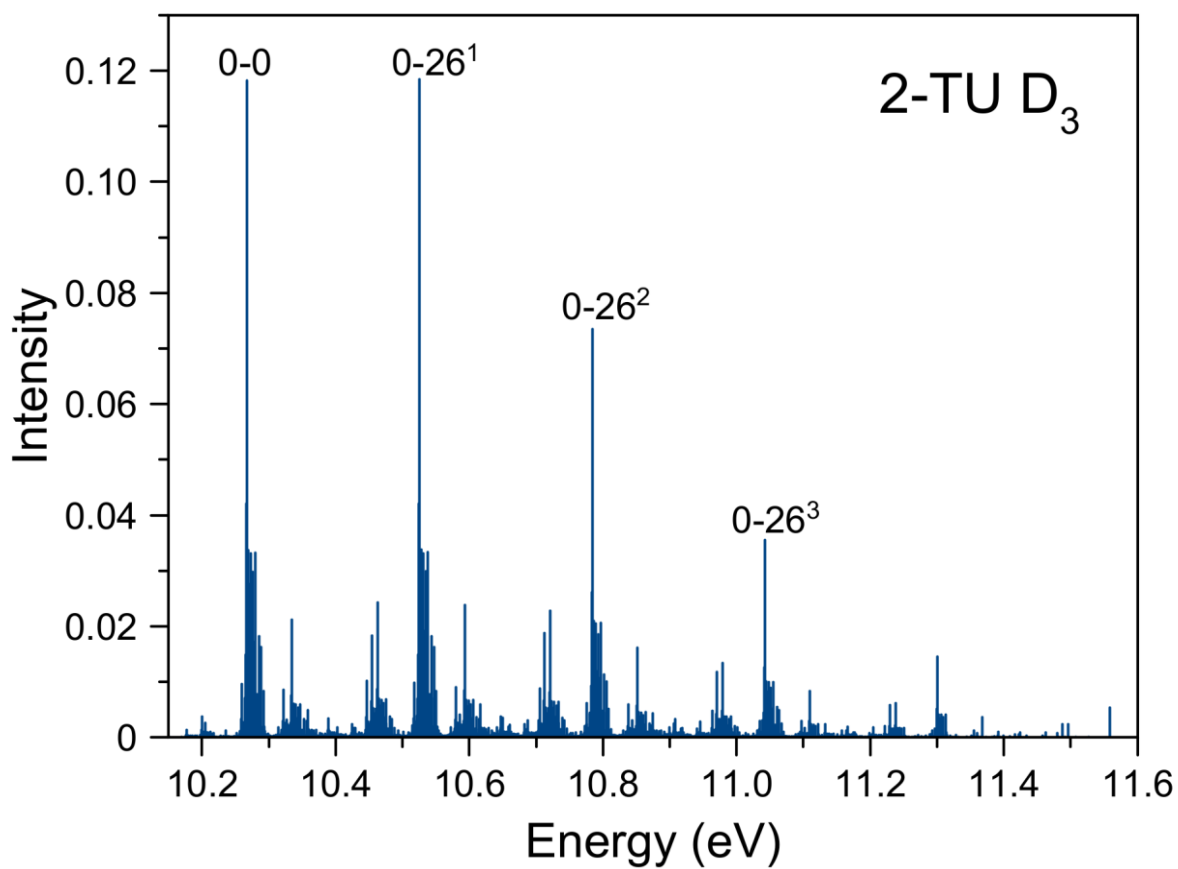

**Figure A5:** $S_0 \rightarrow D_3$ stick spectrum of 2-TU with assignment of major transitions. The used notation is $g^m\text{-}e^n$, where $g$ is the excited normal mode in the $S_0$ state, $e$ is the excited normal mode in the $D_3$ state, and $m$ and $n$ are the numbers of vibrational quanta. "0" stands for the lowest vibrational states of $S_0$ and $D_3$.

**Table A5:** The $D_3$ normal modes of 2-TU involved in the major $S_0 \rightarrow D_3$ vibronic transitions.

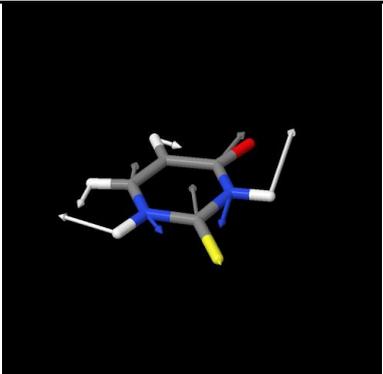

| 2-TU $D_3$ | | |
|---|---|---|
| mode 26, 2083 cm$^{-1}$ | | |

4-TU

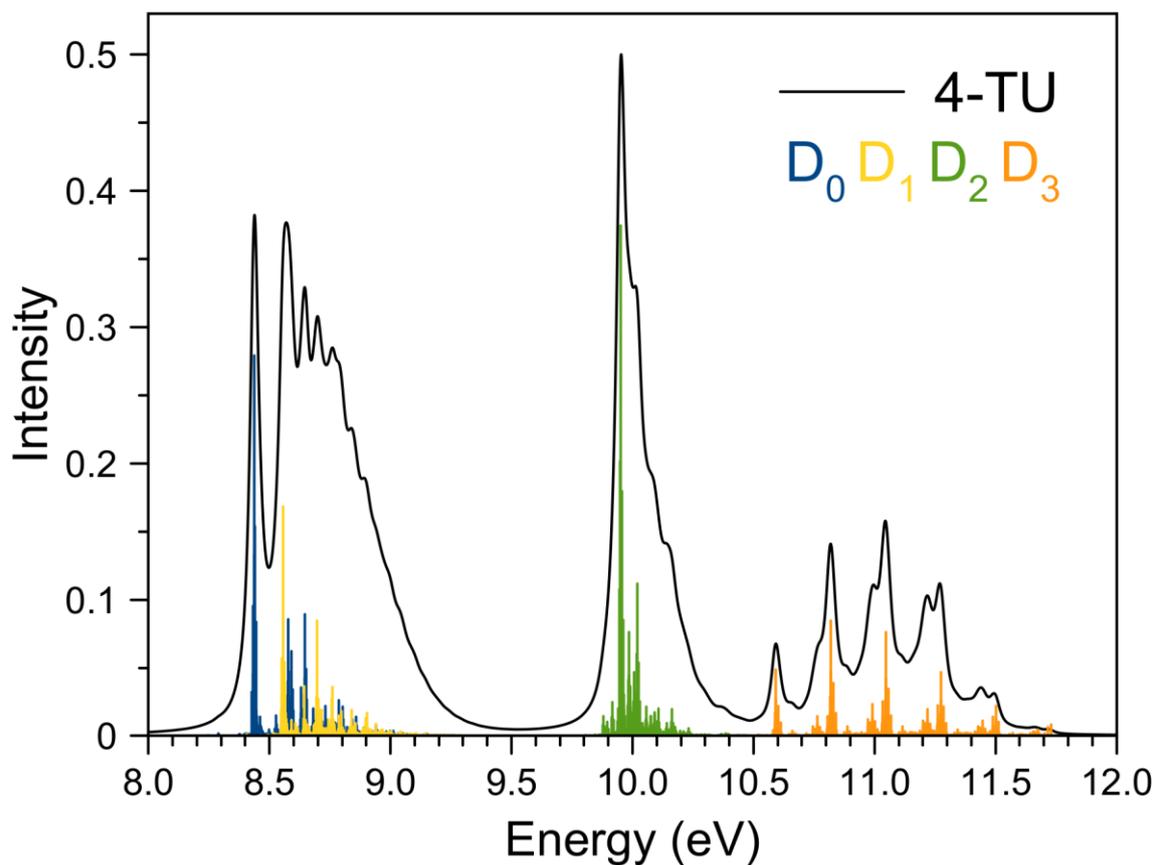

**Figure A6:** Vibronic stick spectra for the lowest four transitions (from $S_0$ to $D_0$–$D_3$) and the broadened spectrum for 4-TU.

**Table A6:** The $S_0$ normal modes of 4-TU involved in the major vibronic transitions from $S_0$ to $D_0$–$D_3$.

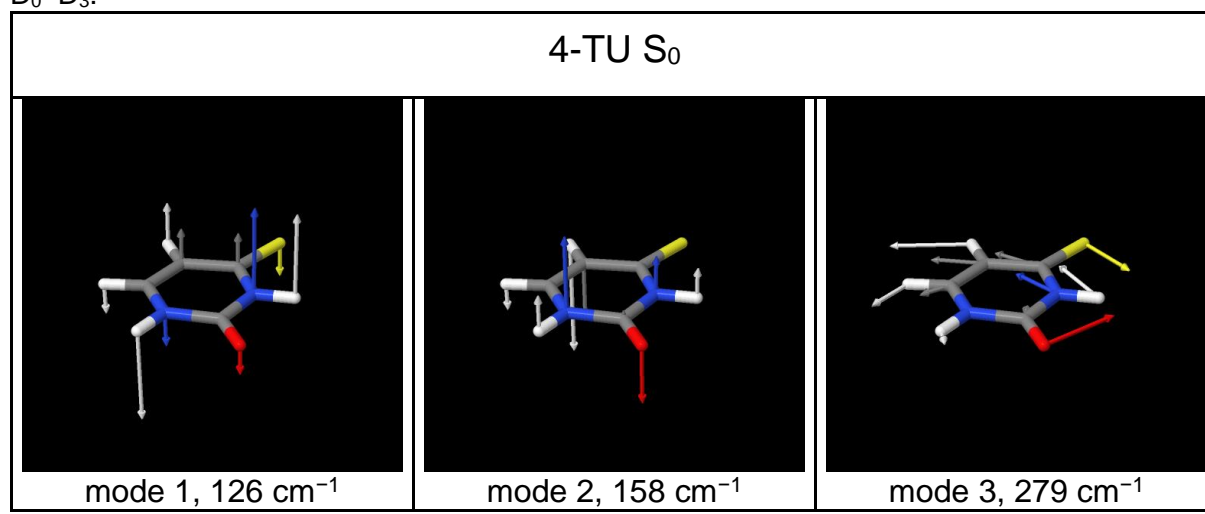

$S_0 \rightarrow D_0$

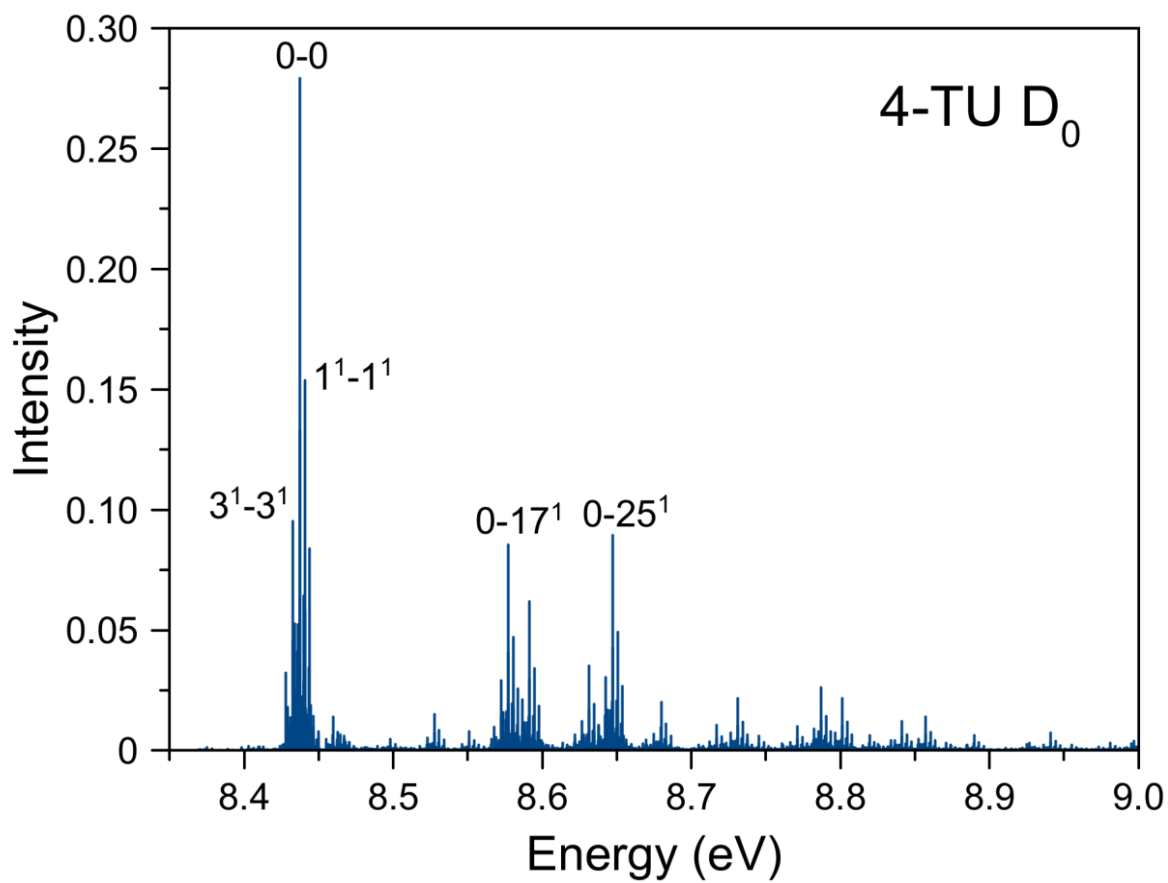

**Figure A7:** $S_0 \rightarrow D_0$ stick spectrum of 4-TU with assignment of major transitions. The used notation is $g^m$-$e^n$, where $g$ is the excited normal mode in the $S_0$ state, $e$ is the excited normal mode in the $D_0$ state, and $m$ and $n$ are the numbers of vibrational quanta. "0" stands for the lowest vibrational states of $S_0$ and $D_0$.

**Table A7:** The $D_0$ normal modes of 4-TU involved in the major $S_0 \rightarrow D_0$ vibronic transitions.

| 4-TU $D_0$ | | |
|---|---|---|
| 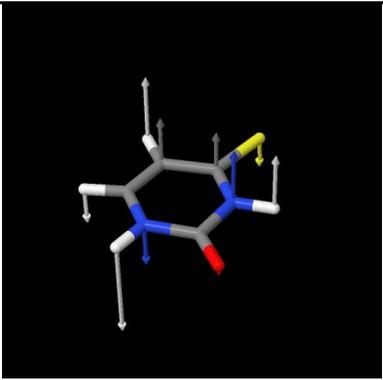 mode 1, 153 cm$^{-1}$ | 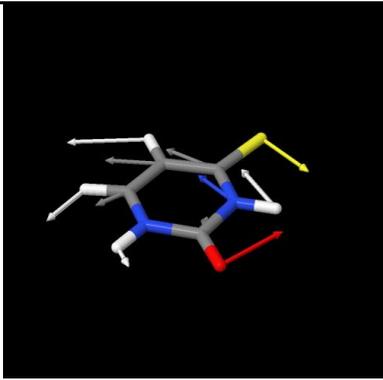 mode 3, 240 cm$^{-1}$ | 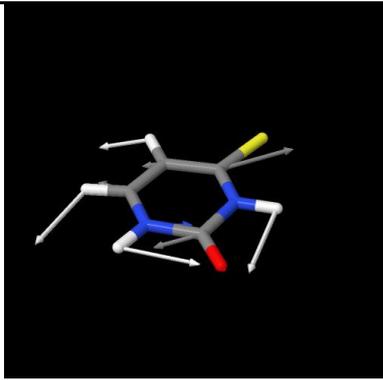 mode 17, 1128 cm$^{-1}$ |
| 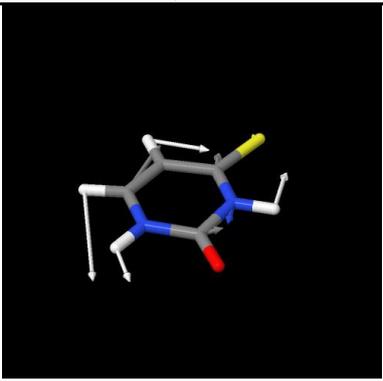 mode 25, 1692 cm$^{-1}$ | | |

$S_0 \rightarrow D_1$

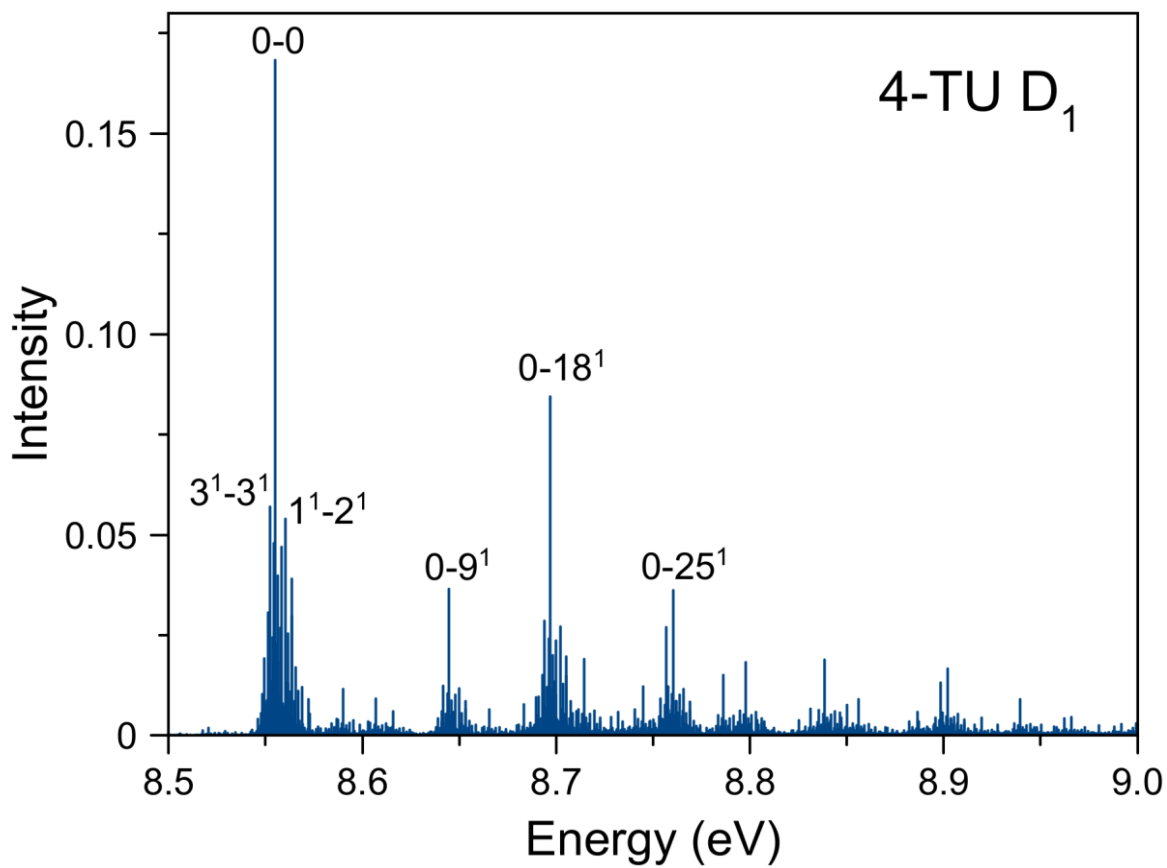

**Figure A8:** $S_0 \rightarrow D_1$ stick spectrum of 4-TU with assignment of major transitions. The used notation is $g^m$-$e^n$, where $g$ is the excited normal mode in the $S_0$ state, $e$ is the excited normal mode in the $D_1$ state, and $m$ and $n$ are the numbers of vibrational quanta. "0" stands for the lowest vibrational states of $S_0$ and $D_1$.

**Table A8:** The $D_1$ normal modes of 4-TU involved in the major $S_0 \rightarrow D_1$ vibronic transitions.

| 4-TU $D_1$ | | |
|---|---|---|
| 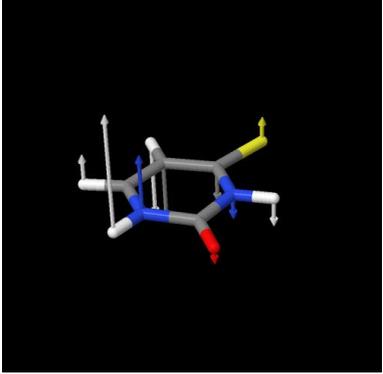 mode 2, 169 cm$^{-1}$ | 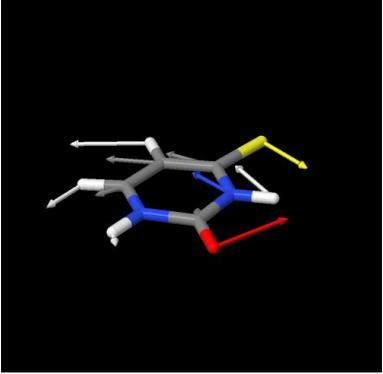 mode 3, 256 cm$^{-1}$ | 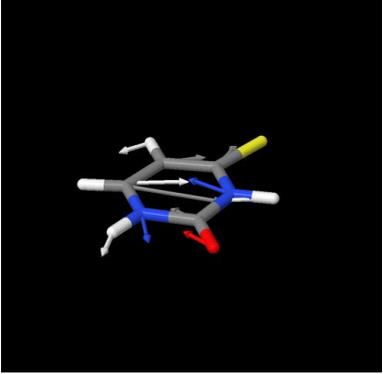 mode 9, 722 cm$^{-1}$ |
| 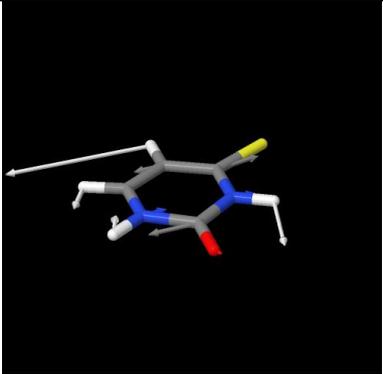 mode 18, 1143 cm$^{-1}$ | 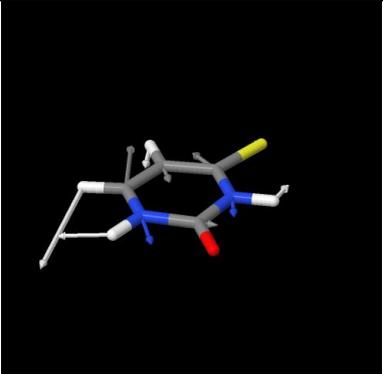 mode 25, 1657 cm$^{-1}$ | |

S₀→D₂

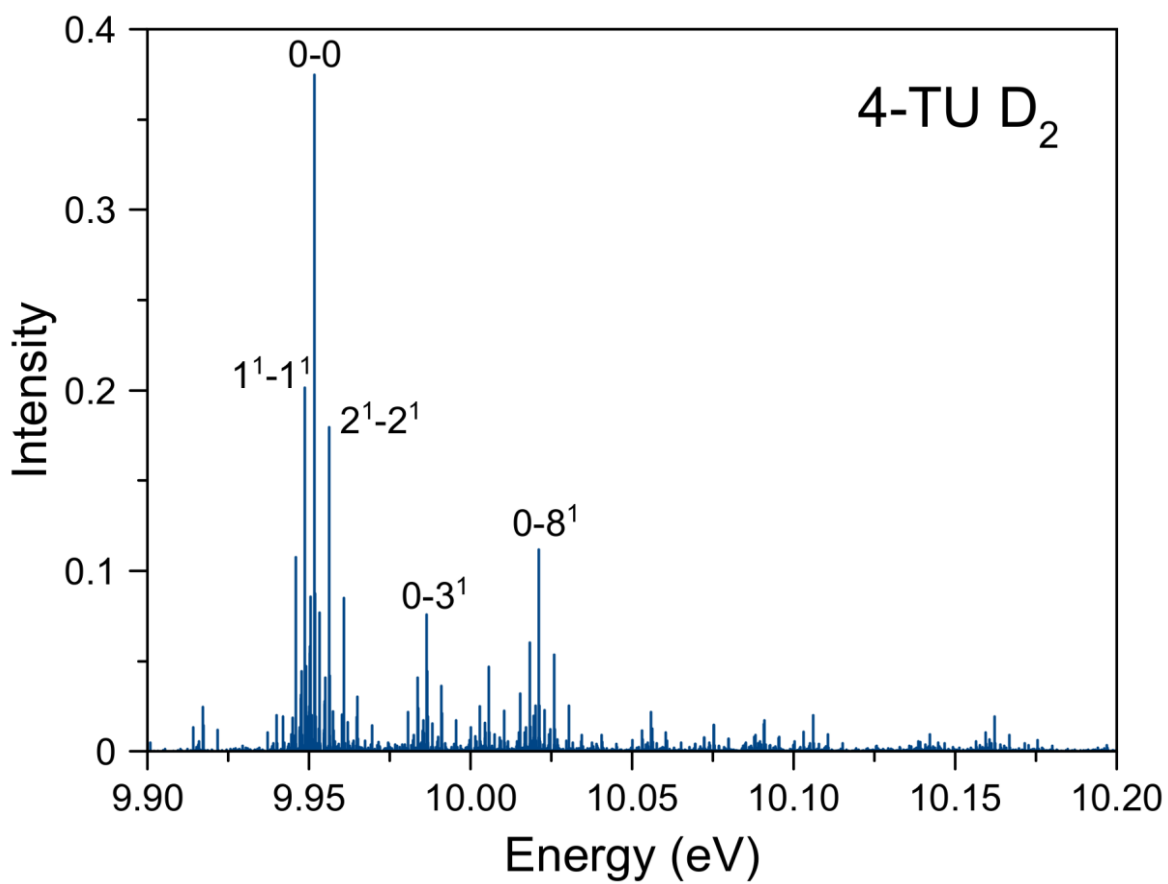

**Figure A9:** $S_0 \rightarrow D_2$ stick spectrum of 4-TU with assignment of major transitions. The used notation is $g^m$-$e^n$, where $g$ is the excited normal mode in the $S_0$ state, $e$ is the excited normal mode in the $D_2$ state, and $m$ and $n$ are the numbers of vibrational quanta. "0" stands for the lowest vibrational states of $S_0$ and $D_2$.

**Table A9:** The $D_2$ normal modes of 4-TU involved in the major $S_0 \rightarrow D_2$ vibronic transitions.

| 4-TU $D_2$ | | |
|---|---|---|
| 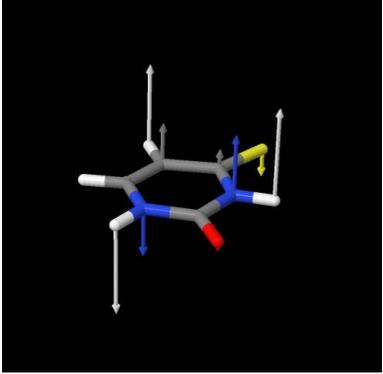 mode 1, 103 cm$^{-1}$ | 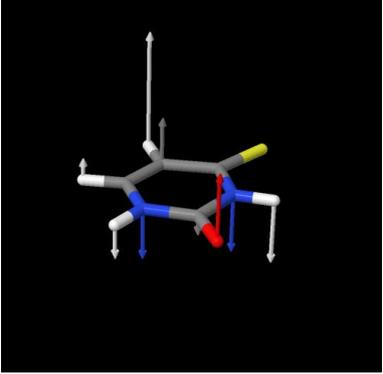 mode 2, 195 cm$^{-1}$ | 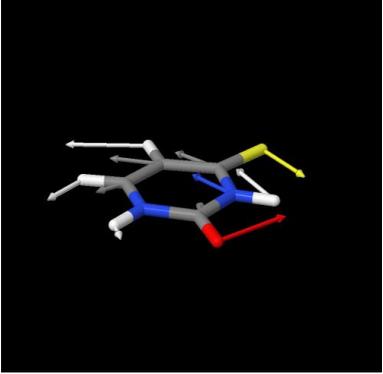 mode 3, 281 cm$^{-1}$ |
| 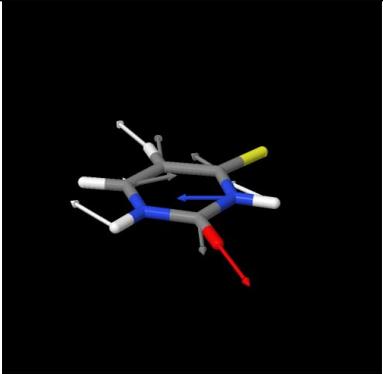 mode 8, 561 cm$^{-1}$ | | |

$S_0 \rightarrow D_3$

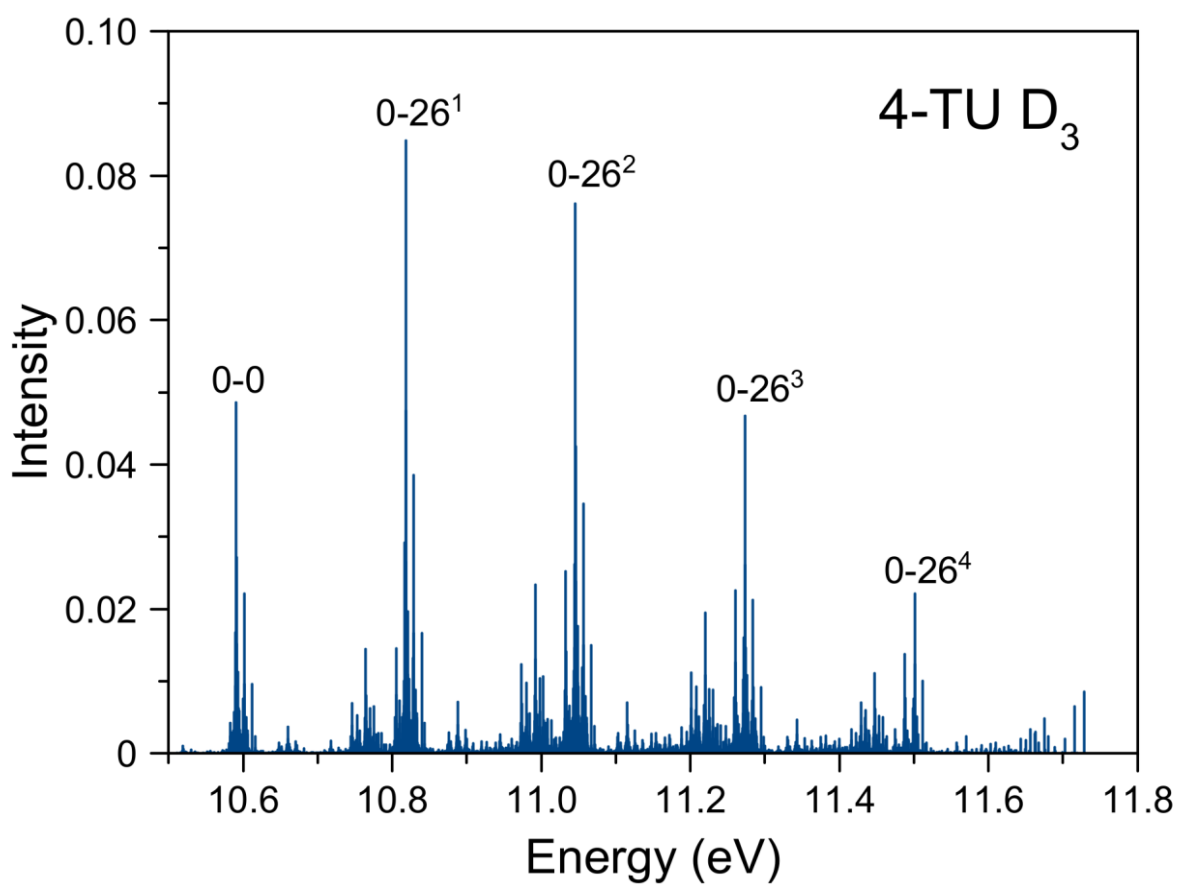

**Figure A10:** $S_0 \rightarrow D_3$ stick spectrum of 4-TU with assignment of major transitions. The used notation is $g^m\text{-}e^n$, where $g$ is the excited normal mode in the $S_0$ state, $e$ is the excited normal mode in the $D_3$ state, and $m$ and $n$ are the numbers of vibrational quanta. "0" stands for the lowest vibrational states of $S_0$ and $D_3$.

**Table A10:** The $D_3$ normal modes of 4-TU involved in the major $S_0 \rightarrow D_3$ vibronic transitions.

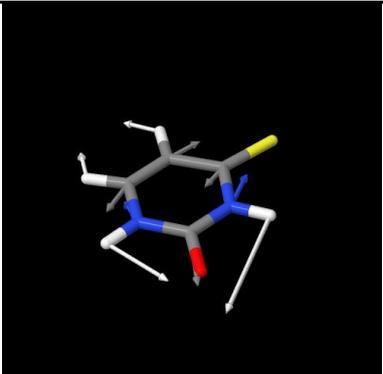

2,4-dTU

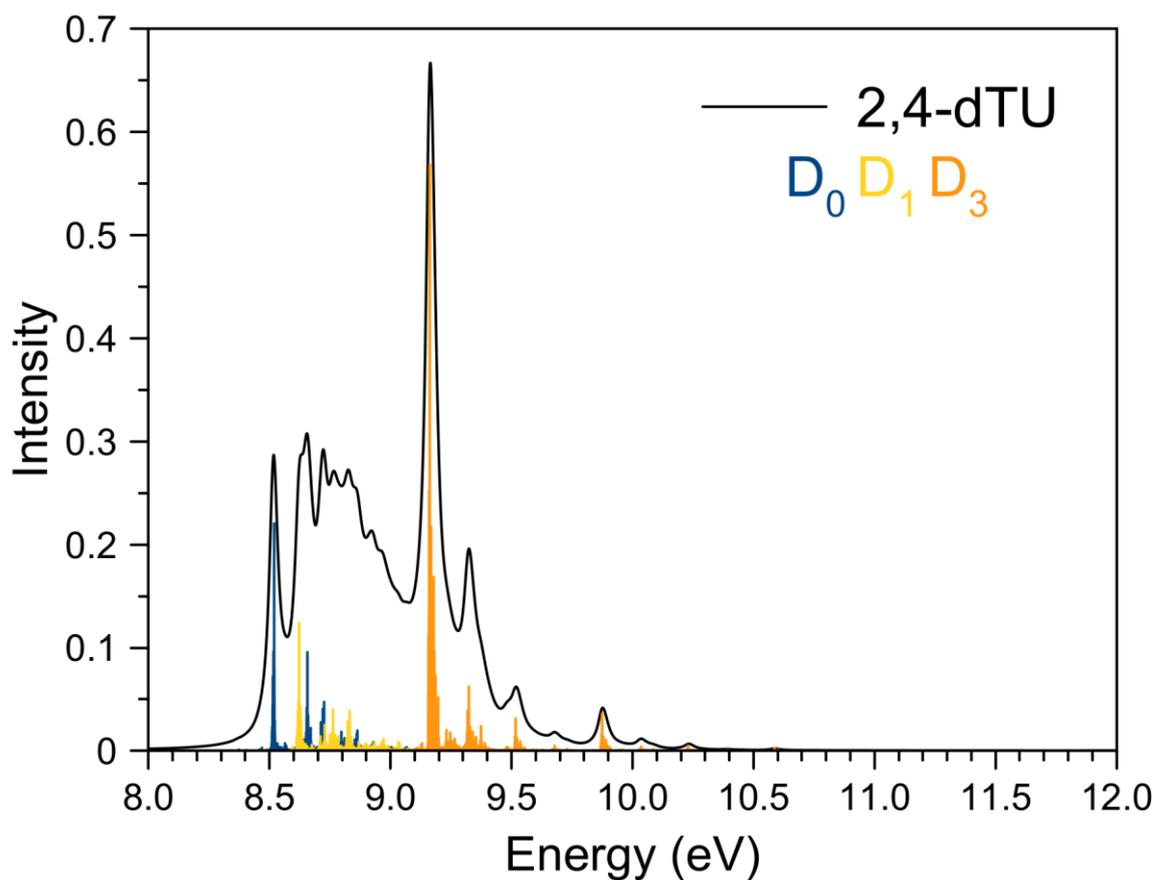

**Figure A11:** Vibronic stick spectra for the transitions from $S_0$ to $D_0$, $D_1$, $D_3$ and the broadened spectrum for 2,4-dTU.

**Table A11:** The $S_0$ normal modes of 2,4-dTU involved in the major vibronic transitions from $S_0$ to $D_0$, $D_1$, $D_3$.

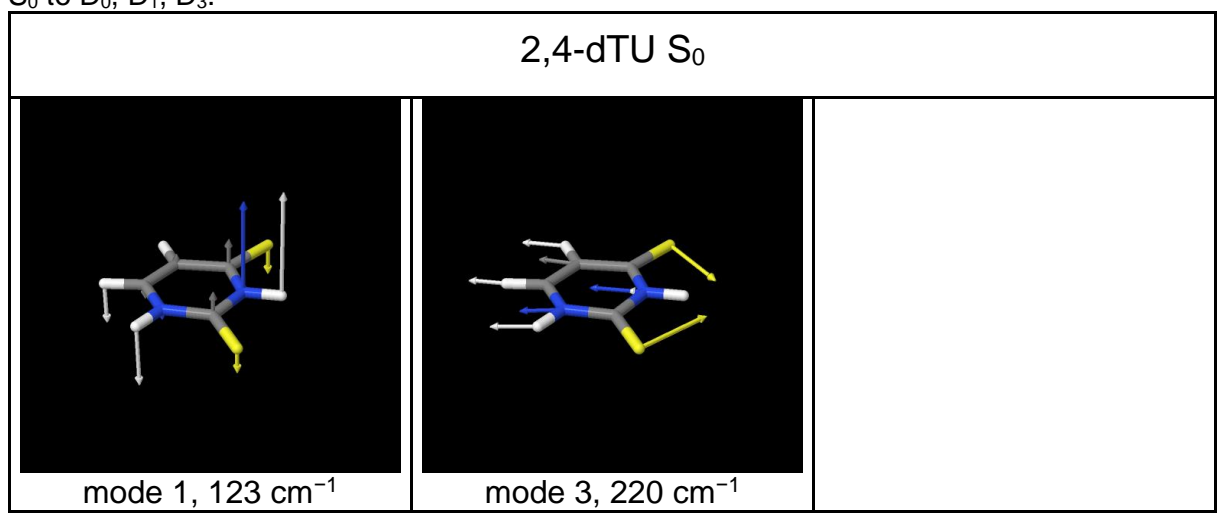

$S_0 \rightarrow D_0$

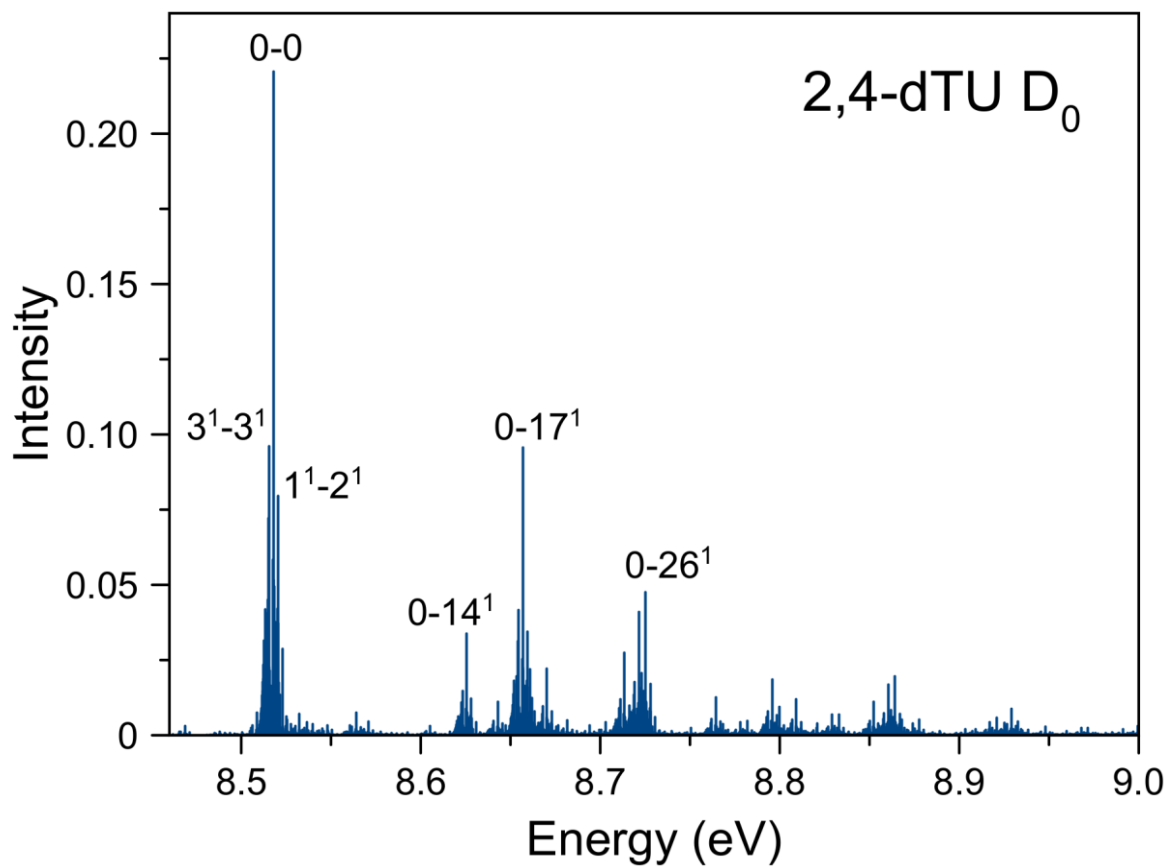

**Figure A12:** $S_0 \rightarrow D_0$ stick spectrum of 2,4-dTU with assignment of major transitions. The used notation is $g^m\text{-}e^n$, where $g$ is the excited normal mode in the $S_0$ state, $e$ is the excited normal mode in the $D_0$ state, and $m$ and $n$ are the numbers of vibrational quanta. "0" stands for the lowest vibrational states of $S_0$ and $D_0$.

**Table A12:** The $D_0$ normal modes of 2,4-dTU involved in the major $S_0 \rightarrow D_0$ vibronic transitions.

| 2,4-dTU $D_0$ | | |
|---|---|---|
| 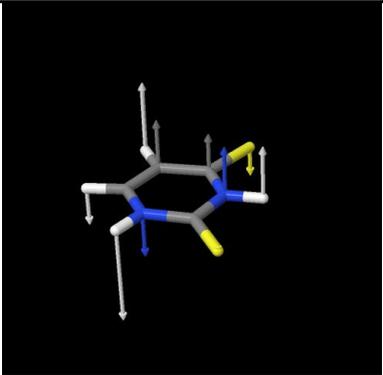 mode 2, 144 cm$^{-1}$ | 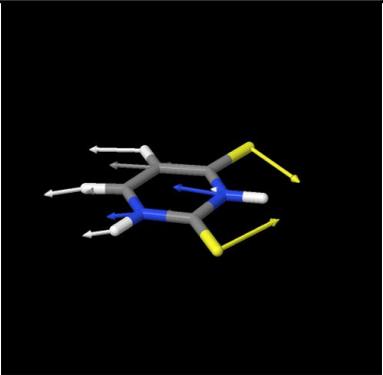 mode 3, 201 cm$^{-1}$ | 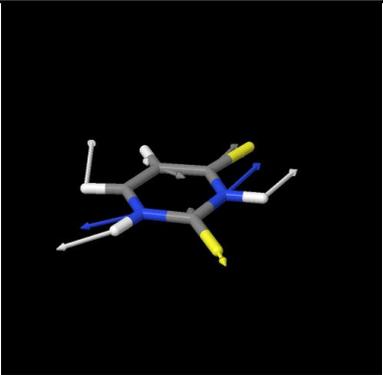 mode 14, 868 cm$^{-1}$ |
| 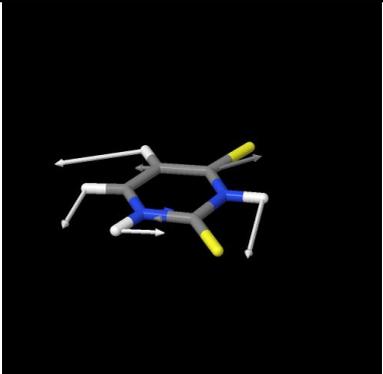 mode 17, 1121 cm$^{-1}$ | 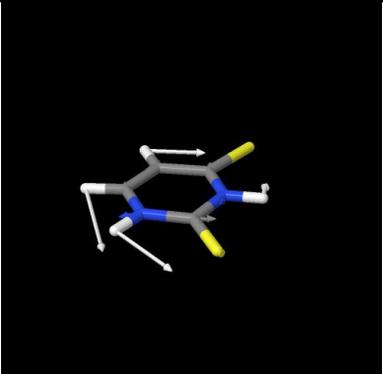 mode 26, 1673 cm$^{-1}$ | |

$S_0 \rightarrow D_1$

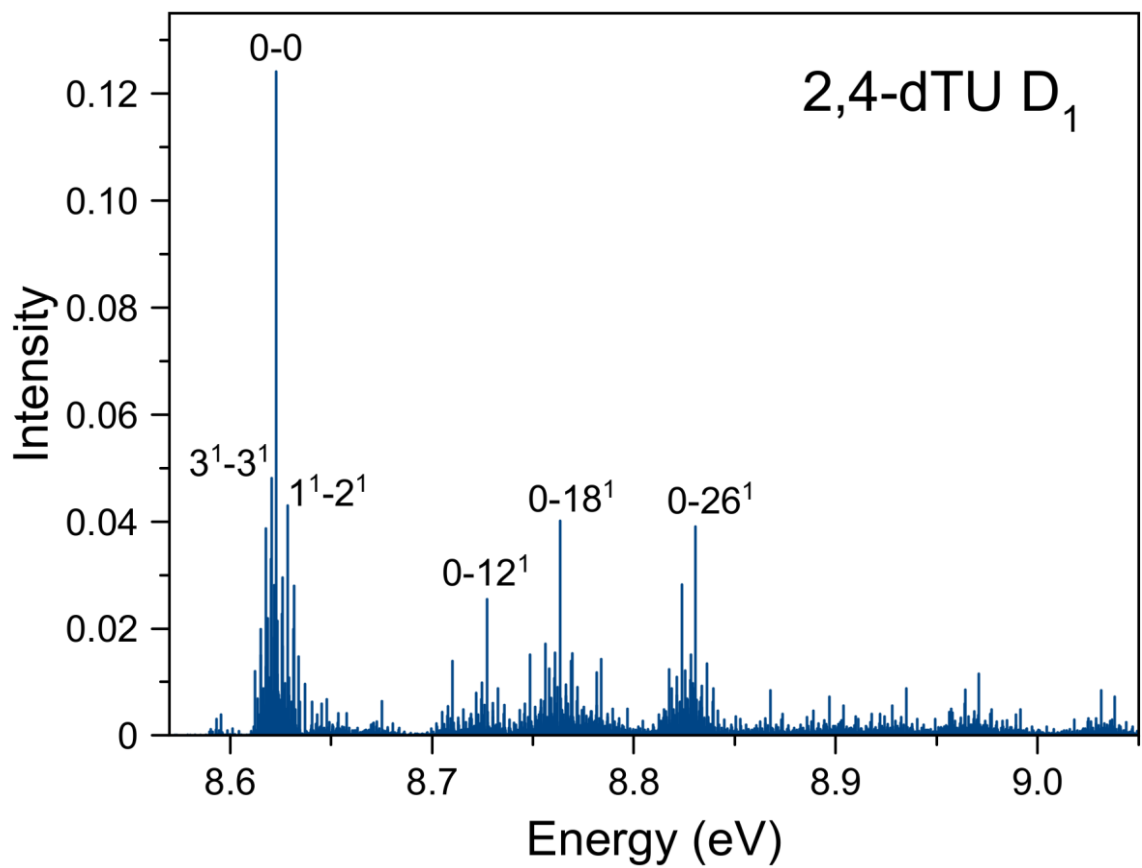

**Figure A13:** $S_0 \rightarrow D_1$ stick spectrum of 2,4-dTU with assignment of major transitions. The used notation is $g^m$-$e^n$, where $g$ is the excited normal mode in the $S_0$ state, $e$ is the excited normal mode in the $D_1$ state, and $m$ and $n$ are the numbers of vibrational quanta. "0" stands for the lowest vibrational states of $S_0$ and $D_1$.

**Table A13:** The $D_1$ normal modes of 2,4-dTU involved in the major $S_0 \rightarrow D_1$ vibronic transitions.

| 2,4-dTU $D_1$ | | |
|---|---|---|
| 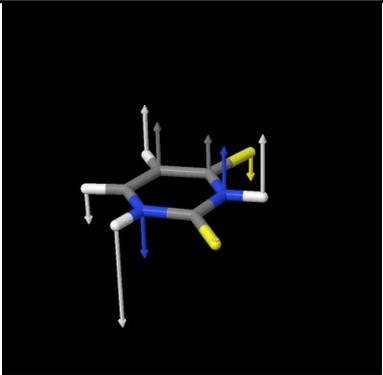 mode 2, 168 cm$^{-1}$ | 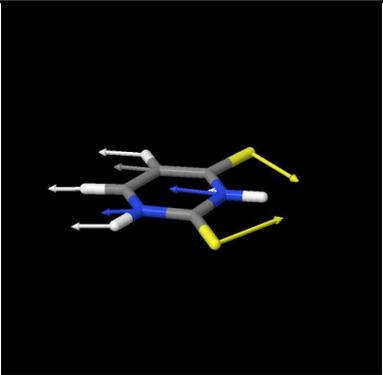 mode 3, 201 cm$^{-1}$ | 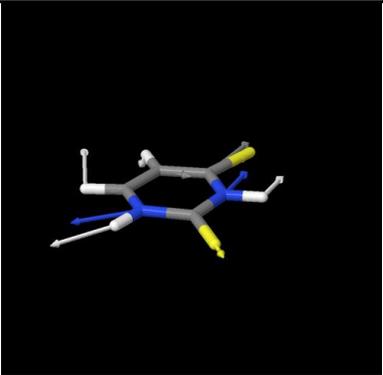 mode 12, 841 cm$^{-1}$ |
| 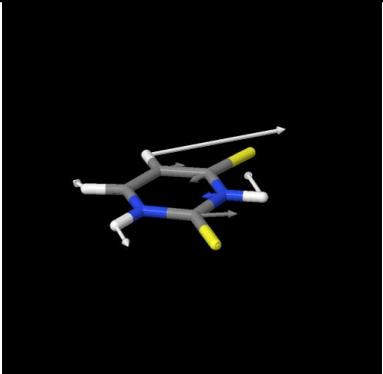 mode 18, 1133 cm$^{-1}$ | 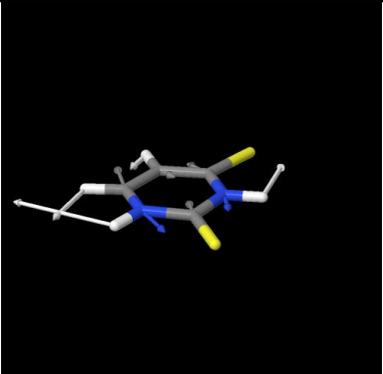 mode 26, 1676 cm$^{-1}$ | |

S$_0$→D$_3$

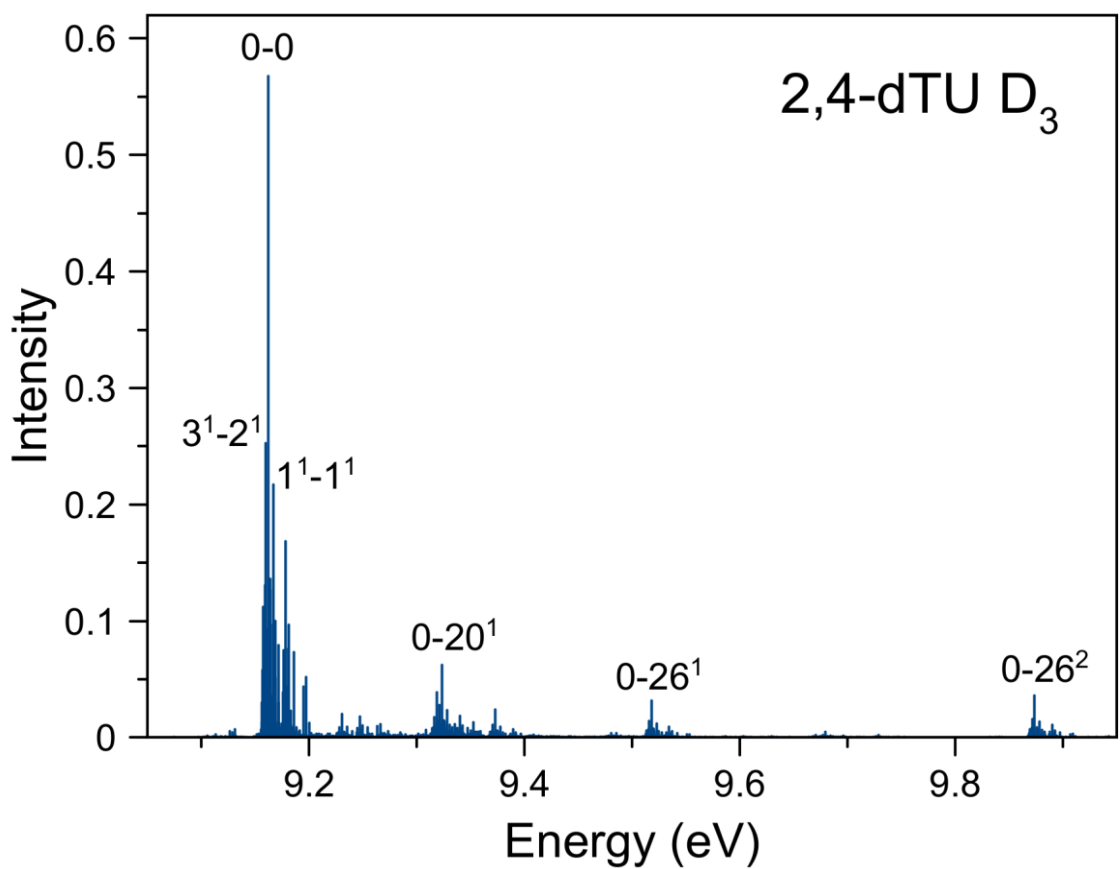

**Figure A14:** S$_0$→D$_3$ stick spectrum of 2,4-dTU with assignment of major transitions. The used notation is $g^m$-$e^n$, where $g$ is the excited normal mode in the S$_0$ state, $e$ is the excited normal mode in the D$_3$ state, and $m$ and $n$ are the numbers of vibrational quanta. "0" stands for the lowest vibrational states of S$_0$ and D$_3$.

**Table A14:** The $D_3$ normal modes of 2,4-dTU involved in the major $S_0 \rightarrow D_3$ vibronic transitions.

| 2,4-dTU $D_3$ | | |
|---|---|---|
| 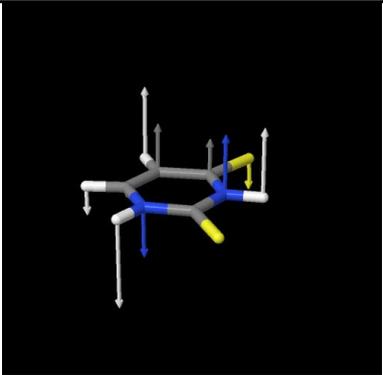 mode 1, 162 cm$^{-1}$ | 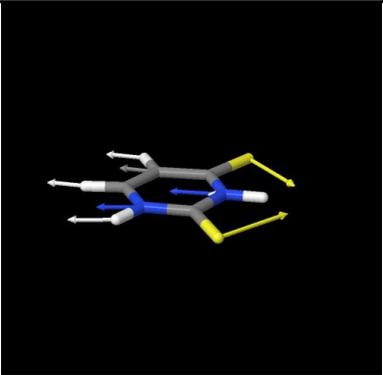 mode 2, 202 cm$^{-1}$ | 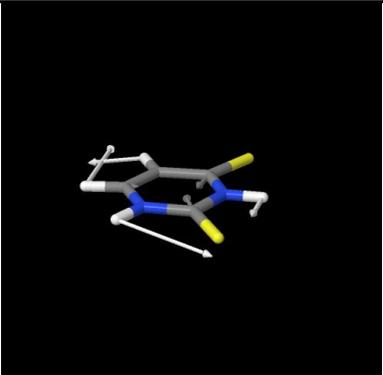 mode 20, 1304 cm$^{-1}$ |
| 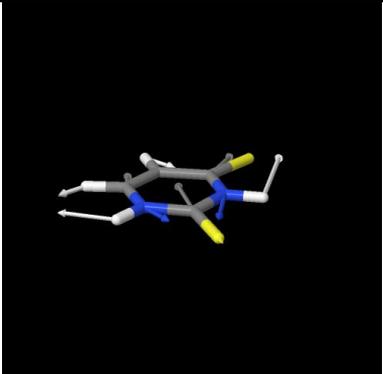 mode 26, 2872 cm$^{-1}$ | | |

# Appendix B. Calculated weak transitions

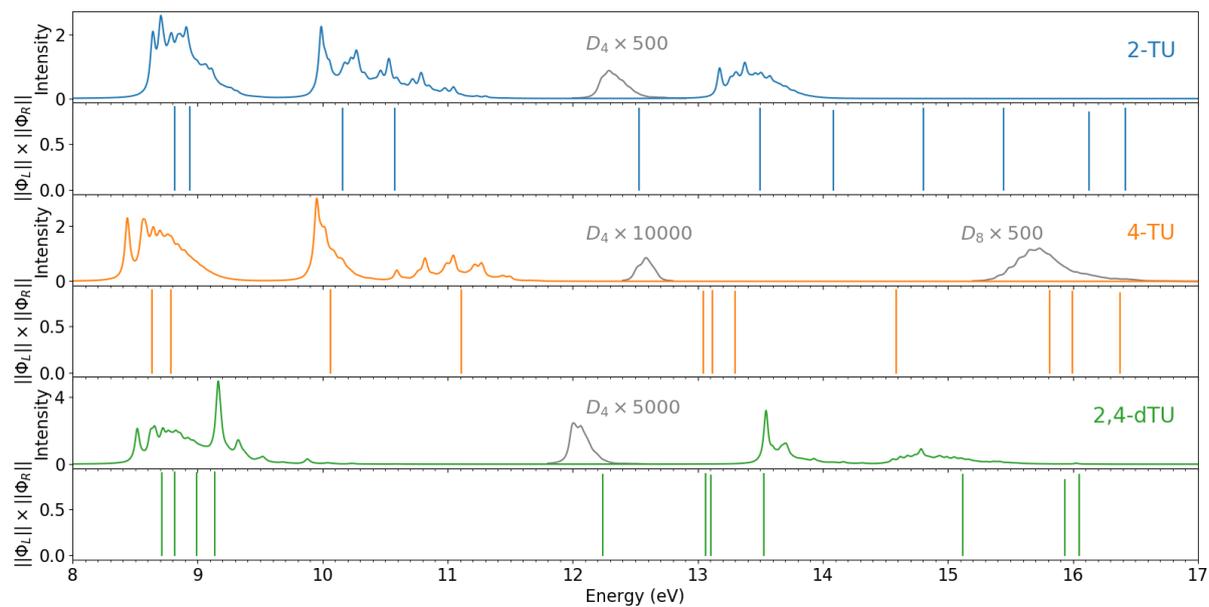

**Figure B1:** The calculated spectra with amplified weak transitions shown in grey.